\documentclass{article}
\usepackage[left=3cm,right=3cm, top=3cm,bottom=3cm]{geometry}
\usepackage[T2A]{fontenc}
\usepackage[utf8]{inputenc}
\usepackage{authblk}
\usepackage[english]{babel}
\usepackage{amsmath,amssymb,amsthm,mathtools}
\usepackage{euscript,mathrsfs}
\usepackage{icomma}
\usepackage{graphicx}
\usepackage{wrapfig}
\usepackage[dvipsnames]{xcolor}
\usepackage{indentfirst}
\usepackage{amsbsy}
\usepackage{wasysym}
\usepackage{cancel}
\usepackage{verbatim}
\usepackage{empheq}
\usepackage{adjustbox}
\usepackage{cite}
\usepackage[mathlines,displaymath]{lineno}
\usepackage[pdfencoding=auto]{hyperref}

\definecolor{urlcolor}{HTML}{120099}
\definecolor{linkcolor}{HTML}{005F5F}
\hypersetup{pdfstartview=FitH, linkcolor=linkcolor,urlcolor=urlcolor, colorlinks=true,citecolor=blue}

\usepackage[framemethod=tikz]{mdframed}
\newmdenv[innerlinewidth=0.5pt, roundcorner=4pt,linecolor=blue,innerleftmargin=6pt,
innerrightmargin=6pt,innertopmargin=6pt,innerbottommargin=6pt]{mybox}

\newcommand{\vphi}{\varphi}
\DeclareMathOperator{\Tr}{\mathrm{Tr}\,}

\newcommand{\bra}[1]{\ensuremath{\left\langle#1\right|}}
\newcommand{\ket}[1]{\ensuremath{\left|#1\right\rangle}}
\newcommand{\brakett}[2]{\ensuremath{\left\langle {#1} | {#2}\right\rangle}}
\renewcommand{\vec}[1]{{\bf #1}}
\newcommand\mean[1]{\ensuremath{\left\langle#1\right\rangle}}

\newcommand\lrb[1]{\left[#1\right]}

\newcommand{\lra}{\quad \Leftrightarrow \quad}

\usepackage{tikz}
\usepackage{subcaption}

\begin{document}

\title{Anatomy of the fragmented Hilbert space: eigenvalue tunneling, quantum scars and localization in the perturbed random regular graph}

\author[a,e]{Daniil Kochergin} 
\author[c,d]{Ivan M. Khaymovich}
\author[f,e]{Olga Valba}
\author[b,e]{Alexander Gorsky}

\affil[a]{Moscow Institute of Physics and Technology, Dolgoprudny 141700, Russia}
\affil[b]{Institute for Information Transmission Problems, Moscow 127994, Russia} 
\affil[c] {Nordita, Stockholm University and KTH Royal Institute of Technology Hannes Alfv\'ens v\"ag 12, SE-106 91 Stockholm, Sweden}
\affil[d]{Institute for Physics of Microstructures, Russian Academy of Sciences, 603950 Nizhny Novgorod, GSP-105, Russia}
\affil[e]{Laboratory of Complex Networks, Brain and Consciousness Research Center, Moscow, Russia}
\affil[f] {Higher School of Economics, Moscow, Russia}

\date{\today}%

\maketitle
\begin{abstract}
 We consider the properties of the random regular graph with node degree $d$ perturbed by chemical
 potentials $\mu_k$ for a number of short $k$-cycles. We analyze both numerically and analytically
 the phase diagram of the model in the $(\mu_k,d)$ plane. The critical 
 curve separating the homogeneous and clusterized phases is found 
 and it is demonstrated that the clusterized phase itself generically is separated 
 as the function of $d$ into the phase with ideal clusters and phase with coupled ones when the continuous spectrum gets formed. The eigenstate spatial structure of the model is investigated and it is found that there are localized scar-like states in the delocalized part of the spectrum, that are related to the topologically equivalent nodes in the graph. 
 We also reconsider the localization of the states in the non-perturbative band formed by eigenvalue instantons and 
 find the semi-Poisson level spacing distribution. The Anderson transition for the case of combined ($k$-cycle) structural and diagonal (Anderson) disorders is investigated. 
 It is found that the critical diagonal disorder gets reduced sharply at the clusterization phase transition, but does it unevenly in non-perturbative and mid-spectrum bands, due to the scars, present in the latter. The applications of our findings to $2$d quantum gravity are discussed.

\end{abstract}
\section{Introduction}\label{Sec:intro}
The interplay between the ergodicity and integrability as well as localization and delocalization
is an interesting problem. The initial one-particle localization in the disordered system
is the well-known starting point~\cite{anderson1958absence}. However it has been recognized more recently
that at least two new generic patterns are possible -- delocalized non-ergodic phase~\cite{Luca2014,Kravtsov_NJP2015} and 
many-body localized phase~\cite{BAA2006,Gornyi2005MBL}. In both cases there are specific mechanisms of the ergodicity
breaking. 

The delocalized non-ergodic phase has been found first in the version of Rosenzweig-Porter model~\cite{Kravtsov_NJP2015} and manifests the fractality of the corresponding wave functions. The underlying origin of the eigenstate fractality and the very non-ergodic phase is attributed to the emergent mini-band structure of the spectrum. 

The combination of interaction
and strong enough disorder amplitudes leads to the many-body localization (MBL) phase 
with full ergodicity breaking~\cite{BAA2006,Gornyi2005MBL,ns3,ns4,ns5}.
It is assumed that the many-body localization in the physical space 
gets mapped into the one-particle localization in a Hilbert space
\cite{altshuler1997quasiparticle}. 
Thus, the Anderson model on random regular graph (RRG) serves as the toy 
model for a identification of MBL phase in the from the Hilbert-space perspective~\cite{Biroli2012difference,Luca2014}. There are many works, devoted to the Anderson model on RRG, showing that the MBL and RRG problems share similarities in their static ergodicity-breaking behavior~\cite{Altshuler2016nonergodic,Altshuler2016multifractal,Kravtsov2018nonergodic,Garcia-Mata2017scaling,Parisi2019anderson,LN-RP_RRG,Tikhonov2016fractality,Tikhonov2017multifractality,avetisov2016eigenvalue,avetisov2020localization,valba2021interacting,Garcia-Mata2020two},
dynamical~\cite{Biroli2017delocalized,bera2018return,DeTomasi2019subdiffusion,Colmenarez2022subdiffusive,Biroli2020anomalous,Tikhonov2019statistics,Tikhonov2021eigenstate,LN-RP_dyn} properties, as well as finite-size corrections~\cite{Tikhonov2016anderson,Biroli2018delocalization,LN-RP_WE}, see~\cite{Tikhonov2021from} for the recent review. 
Thus, a better understanding of the RRG-like models is relevant for the MBL problem.

Therefore the investigation of the Anderson localization per se and other
possible non-trivial phases in perturbed RRG is an interesting issue. 
In particular, RRG provides the suitable playground for investigation
the effect of the structural disorder on the localization. To this 
aim the chemical potentials for the short cycles can be added and 
the clusterization phase transition takes place at some critical values
of the chemical potentials. The clusterization amounts to the controlled
structural disorder for the Anderson problem on the graph. 

Note that the exponential random graphs, including RRG, are the 
statistical models that are the discrete versions of the matrix models,
when instead of the generic elements of the matrix there are the bimodal
elements of the adjacency matrix of the graph. The clusterization phase
transition in the exponential random graphs corresponds to the condensation of eigenvalue
instantons, individual eigenvalue instanton corresponds exactly to a emerging cluster~\cite{nadakuditi2013spectra}. The effects of the eigenvalue instantons
in the matrix models has been discussed in the several contexts and they
are related to the creation of the baby Universe in $2$d gravity~\cite{shenker1991strength,david1993non},
ZZ branes in the Liouville theory and to the partial violation of the
gauge theory in the super-symmetric (SUSY) gauge theories, see~\cite{2014ForPh..62..455M} for the review. 
However, in the matrix models only the large $N$ planar limit is analytically tractable
and all non-perturbative instanton effects are suppressed by $\exp(-N)$ factor.
On the other hand, in the RRG ensemble we can model the non-perturbative effects numerically
at finite $N$ in a reasonably simple manner. Note also that quite recently the new type
of the eigenvalue hole-like instantons have been identified~\cite{marino2022new}
and we shall argue that there is its clear-cut RRG counterpart. 

The underlying mechanisms and examples of one-particle localization in the Hilbert space are 
interesting in a context of the MBL phase of interacting many-body system in the physical space. 
The Hilbert-space fragmentation is one of the mechanisms, considered as the origin
of MBL phase. Typically it is attributed to the hidden symmetry of the 
many-body systems or emerging local conservation laws see~\cite{pakrouski2020many,ren2021quasisymmetry,pakrouski2021group,o2020tunnels,sun2022majorana}. 
The formation of the quantum many-body scars (QMBS) is another mechanism for localization 
in the Fock space, see~\cite{moudgalya2021quantum,chandran2022quantum,serbyn2021quantum} for
reviews. The QMBS represents the non-thermalizing energy state, protected by some emerging symmetry. 
There are two types of QMBS states: one family forms the equally spaced levels and can be related
to some representation of the underlying symmetry group. The second class corresponds to the isolated QMBS mid-spectrum states, typically at zero energy $E=0$. There can be a few or exponentially many of such states. The most common mechanism behind an emergence of such states in the mid-spectrum has been suggested in~\cite{shiraishi2017systematic}, however it is not the only one,
see, e.g.,~\cite{banerjee2021quantum,iadecola2019exact,jeyaretnam2021quantum,moudgalya2022hilbert,ok2019topological,srivatsa2020quantum,mondaini2018comment,khemani2020localization}.

In some cases the Hamiltonian of the many-body system can be mapped into the adjacency matrix of the graph or ensemble of graphs, the PXP is the simplest example of such situation.
In this case the spectral theory of graphs enters the game of scar hunting.
The exact scars have been found for $E=0$, $E=\pm \sqrt{2}$ in PXP model~\cite{lin2019exact} and at 
$E=\pm \sqrt{n}$ in more general models belonging to this family~\cite{surace2021quantum,surace2021exact,2023arXiv230508123J}. In~\cite{surace2021exact}
the ensemble of graphs with some parameter, responsible for 
the probability to have a link, has been considered and it has been argued that there is a transition in the parameter space between the phase with
negligible amount of scars and the scarred phase. The very presence of scars
has been attributed to the probability to realized particular subgraphs in the ensemble.

In this paper, we investigate the localization properties in RRG networks,
perturbed by the chemical potentials for the number of short cycles~\cite{avetisov2016eigenvalue,avetisov2020localization,valba2021interacting,kelly2019self}. These systems can be
considered as the models with large number of conservation laws - the degree conservation 
for each node plays such role. The model undergoes the clusterization phase transition
at the critical chemical potentials for $3$-cycles~\cite{avetisov2016eigenvalue} and $4$-cycles~\cite{valba2021interacting,kelly2019self}, similar to that of the Hilbert-space fragmentation. It was found that the
spectrum of the graph Laplacian above the phase transition involves the 
non-perturbative (side) band, where the number of the soft modes for each realization
equals to the number of clusters as well as the continuum band. The localization properties of the clustered phase have been investigated before in~\cite{avetisov2020localization} . 

The density of states (DOS) in the perturbative (mid-spectrum) band for high chemical potential of $3$-cycles $\mu_3>\mu_c(3)$ and $d=20$ does not fit with the Wigner semicircle
and has the triangle shape typical to the scale-free networks~\cite{goh2001spectra}. Moreover it has been checked 
in~\cite{avetisov2020localization} that upon the randomization procedure which respects the degree conservation 
and the fixed number of clusters the triangle shape gets transformed into the Wigner semicircle. 
This means that the clustered phase has a kind of the memory about the initial state and the states in perturbative (mid-spectrum) band have to be considered as a sort of non-ergodic. 

\begin{figure}[p]
\centering
\includegraphics[width=\textwidth]{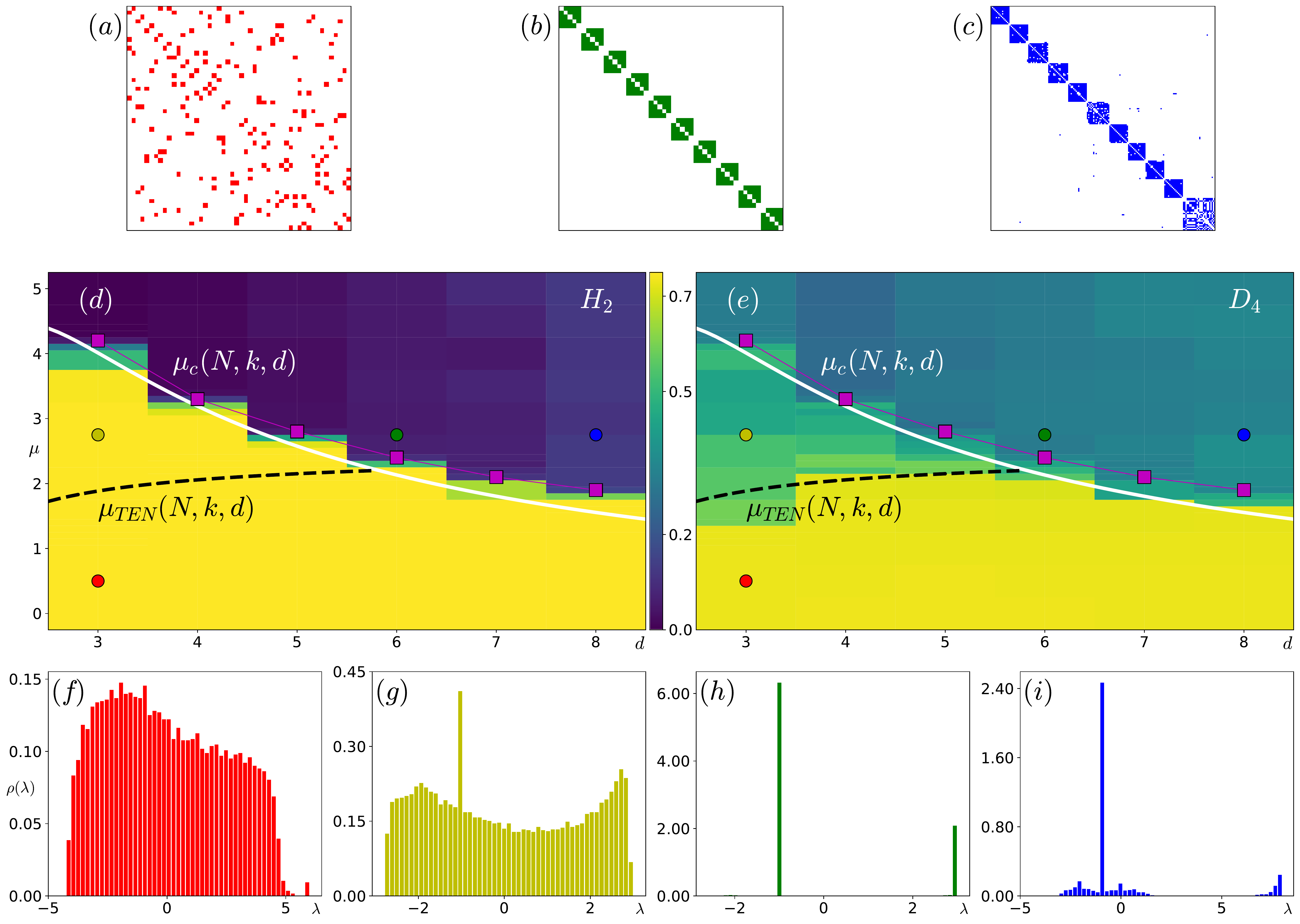}
\caption{Phase diagram in the plane ``chemical potential~--~vertex degree'' $(\mu_k,d)$ for finite-size RRG graphs for $N=256$ and $k=3$-cycles. (a-c)~The cluster structure of RRG in (a)~unclustered; (b)~ideally clustered, and (c)~interacting clustered phases. (d-e)~Phase diagram, with drastic changes (d)~in the density of states (DOS) via the Hellinger distance with respect to the ideal cluster, showing the clusterization transition (purple squares), and (e)~in the higher-order fractal dimension $D_4$, sensitive to the scar states, given by the topologically equivalent nodes (TEN).
Panels~(f-i) show the averaged DOS in each of the $4$ phases:
(f)~unclustered, (g)~TEN-scarred unclustered, (h)~ideally clustered, and (i)~interacting clustered phases. 
The colors of the solid circles in the panels (d, e), marking each of $4$ phases, correspond to the colors of the blocks in (a-c) and the DOS in (f-i). Solid white $\mu_c(N,k,d)$, Eq.~\eqref{eq:mu_c(N,k,d)}, and dashed black $\mu_{TEN}(N,k,d)$, Eq.~\eqref{eq:mu_TEN}, lines show analytical estimates for the transition lines between the above phases.
}
\label{fig:fig1_phase_diag} 
\end{figure}

In this study, we reconsider the spectral properties of the perturbed RRG and extend the analysis for the model with additional diagonal disorder. Our findings are as follows
\begin{itemize}
\item Both analytically and numerically, we find the phase diagram on the $(\mu_k,d)$ parameter plane for different $k$-cycles and system sizes $N$ and identify the
critical curves, separating the homogeneous, scarred and two clusterized phases (see Fig.~\ref{fig:fig1_phase_diag}).

 \item We find the quantum scar states in the spectrum in both an unclustered phase, below the clusterization transition, 
 and in the regime of non-ideal interacting clusters.
 It turns out that scars get identified as the emerging topologically equivalent nodes (TENs) 
 and are localized at the specific values of energy. 
 
 \item We reconsider the properties of the non-perturbative bands formed by clusters and 
 found that level-spacing distribution corresponds to the semi-Poisson statistics. 
 
 \item The Anderson transition for the case of the combined (structural + diagonal) disorder is analyzed and it is shown that with clusterization the critical disorder drastically reduces, but it does so unevenly for the mid-spectrum band with TENs and the non-perturbative band.
 
\end{itemize}

The quantum scars analogues can be identified in this model as few states, localized on the RRG, on the background of the rest metallic states. One of the ways to determine the former ones is to observe the jumps in the higher-order inverse participation ratio (IPR), $IPR_q, q>2$, averaged over the spectrum, as a function of the $k$-cycle chemical potential $\mu_k$, and these jumps are complemented by the emerging peaks in the density of states at the center of the perturbative zone. 
It turns out that these scar-like states have an interesting origin. There can be the topologically equivalent nodes (TEN) in the graph, the measure of equivalence can be different. The pairs of such TEN form the effective dipoles and the analysis shows that each such pair supports a scar. Higher topologically equivalent multipoles of TEN can be formed as well. 
We argue that the number of TENs grows with the chemical potential $\mu_k$ as we increase the number of $k$-cycles. It is this presence of TENs, which explains the unexpected triangular shape of DOS, observed in the clustered phase in~\cite{avetisov2016eigenvalue}. If we apply the randomization process~\cite{avetisov2020localization} to the cluster structure, the TENs get destroyed. This explains why the Wigner semicircle in the perturbative band gets restored from the triangle at the cluster randomization. 

In such a way, TENs can be considered as the nuclei of the clusterization process and the corresponding peak is clearly seen in DOS before the clusterization transition, see Fig.~\ref{fig:fig1_phase_diag}.
In the paper for strong enough $\mu_k$ chemical potential, we also observe a trivial kind of TENs, being the so-called ideal clusters. 
These are the clusters that involve only $k$-cycles with no lower cycles and they form an ideal-clustered phase at large $\mu_k$ and intermediate node degrees $d$. 
However, the increasing the node degree $d$ at fixed chemical potential $\mu_k$ leads to the interaction between the above ideal clusters and to the formation of the non-ideal clustered phase. The latter phase is also populated by non-trivial TEN-scarred states.


We also analyze the localization of wave functions in the non-perturbative cluster band and extend the analysis of~\cite{avetisov2020localization} to the level spacing distribution. 
It is found that the wave functions develop more complicated localization properties. 
Some analogue of the "particle-hole" structure can be developed. The level statistics 
is found to be semi-Poisson, which is also an attribute of several kinds of Bethe-ansatz integrable models, like the Richardson's model~\cite{Yuzbashyan_NJP2016,Nosov2019mixtures,Motamarri2022RDM}.

The RRG with the diagonal disorder is a benchmark toy model for the Anderson localization in the Fock space~\cite{altshuler1997quasiparticle,Biroli2012difference,Luca2014}. We investigate the combined effect of the diagonal disorder and the Hilbert-space fragmentation, given by the clusterization at large $\mu_k$, on the Anderson phase transition. 
It is found that the clusterization strongly decreases the critical diagonal disorder $W_{cr}$, when the complete localization takes place. We also find that the competition between the TENs and the diagonal disorder, which takes place in the perturbative band, leads to the non-monotonic localization properties.
Initially degenerate TENs, forming a flat-band structure with spread wave functions, first becomes localized at small diagonal disorder, due to lifted degeneracy of their levels.
But, second, as disorder increases, TENs get destroyed and, thus, delocalized, and only at even higher disorder all the states in the perturbative band are Anderson localized.
Absence of TENs in the side non-perturbative band leads to the Anderson localization there at smaller disorder strength, thus, opening the possibility to realize an effective mobility edge.

The paper is organized as follows. In Section~\ref{Sec:Phase_diag} we recall some results concerning the 
clusterization of exponential
random graphs and find the phase diagram of the model. In Section~\ref{Sec:TENs} we investigate the phase diagram, by numerically finding the non-thermalizing quantum scars, and analytically explain their position in the spectrum.
The scars are identified with topologically equivalent nodes. In Section~\ref{Sec:semi-Poisson_side_band} we investigate
in some details the localization properties in the non-perturbative band and then in Sec.~\ref{Sec:ALT} consider
the Anderson transition in the model with combined structural and diagonal disorder. 
We show that there is the sharp decrease of the critical diagonal disorder at the clusterization transition point. The applications of our findings 
to the $2$d quantum gravity are present in Sec.~\ref{Sec:gravity}.
The results and open questions are summarized in Conclusion, Sec.~\ref{Sec:Conclusion}. 
In Appendices we present the plots of $d$-dependence of DOS for the $\mu_k$-perturbed RRG, with $k>3$ and
present the example of complex TEN.

\section{Phase diagram of the model}\label{Sec:Phase_diag}
In this Section, first, we define the model, Sec.~\ref{SubSec:Model}, and consider the phase diagram, extracted from numerics, Sec.~\ref{SubSec:phase_diag}.
Next, we focus on the clusterization part of the phase diagram analytically and consider the structure of the ideal-clustered phase in Sec.~\ref{SubSec:ideal_cluster}, and estimate the critical clusterization line by comparing the pure RRG and ideal-clustered phases in Sec.~\ref{SubSec:mu_c}.

\subsection{Exponential random graphs}\label{SubSec:Model}
Let us recall the Anderson model on exponential random graphs.
We focus at the RRG ensemble, where the degrees of all nodes are fixed to $d$ and the
partition function is considered
\begin{equation}
 Z(\mu_k)= \sum_{RRG} \exp(-\sum_{k} \mu_k M_k ) \ ,
 \label{eq:rrg_mu_weights}
\end{equation}
where $M_k$ is the number of the length-$k$ cycles in the graph without 
the back-tracking and $\mu_k$ are the chemical 
potentials counting the number of these $k$-cycles. The leading contribution
into the $M_k$ comes from the term $TrA^k$ where $A$ is the adjacency matrix of the
graph.
We consider the spinless fermions on the graph and the graph adjacency matrix $A$ serves as the Hamiltonian 
\begin{equation}\label{eq:eigenproblem}
 A\Psi_{\lambda}=\lambda \Psi_{\lambda} \ . 
\end{equation}
Further, we focus in~\eqref{eq:rrg_mu_weights} on the case of only one non-zero (in most of the cases positive) chemical potential $\mu_k$.
In order to sample such $\mu_k$-weighted RRG numerically, we follow the standard procedure of Monte-Carlo annealing~\cite{avetisov2016eigenvalue} from the pure RRG, with all $\mu_k=0$, via random rewiring process. If the auxiliary annealing time is large enough, the resulting RRG samples the distribution, given by~\eqref{eq:rrg_mu_weights}. 

For RRG, the eigenproblem of a Laplacian $L= D-A$ and the above adjacency matrix have common eigenstates and related spectra, since a degree matrix $D = d\cdot I$ is proportional to the identity $I$. 
Therefore here and further we focus on the adjacency matrix problem (unlike mentioned otherwise).

Upon the averaging over the ensemble DOS can be obtained
\begin{equation}
 \rho(\lambda)=\frac{1}{N}\left\langle\sum_i\delta(\lambda-\lambda_i)\right\rangle_{RRG} \ ,
\end{equation}
which in the large $N$ limit at $\mu_k=0$ has the Kesten-McKay form for the adjacency matrix
\begin{equation}
 \rho_{KM}(\lambda)= \frac{d \sqrt{\left[4(d-1)-\lambda^2\right]}}{2 \pi (d^2-\lambda^2)}
\end{equation}

At small chemical potentials the partition function 
is dominated by smooth RRG graphs while above some critical values
of the chemical potentials the clustered RRG phase emerges and the
number of clusters is fixed by the node degree $d$~\cite{avetisov2016eigenvalue, valba2021interacting}. The structure of the
clustered phase depends on the value of $k$ -- for odd cycles we have the 
ordinary clusters while for even $k$ the clusters are bipartite. If the
additional hard-core constraint is imposed for the case $\mu_4 \neq 0$
the bipartite clusters turn out to be hypercubes~\cite{valba2021interacting,kelly2019self}.

The clusterization can be properly probed via the spectral analysis since
it is known from the network theory~\cite{nadakuditi2013spectra} that each eigenvalue escaped
from the continuum corresponds to the cluster in the network. Generically escaped eigenvalues
interact with each other and finally upon the averaging over ensemble 
form the non-perturbative bands in the spectrum of the RRG Laplacian. However,
different patterns of clusterization transition are possible and details depend on the parameters of the model.
Due to the clear-cut spectral identification of the clusters and their controllable number, the RRG ensemble is very suitable for the 
numerical investigation of the Hilbert-space fragmentation at finite system sizes $N$
and, more generally, for the analysis of the non-perturbative phenomena in the matrix models. 

Another general question concerns 
the localization properties 
of the eigenstates of the graph Laplacian in the Anderson model.
The previous studies mainly focused at the dependence of Anderson localization
on RRG on the diagonal disorder, see~\cite{Tikhonov2021from} for the review.
and not much attention was paid for the effects of structural disorder. Some partial results concerning clusterized phase of RRG
at the fixed values of parameters have been reported in~\cite{avetisov2020localization}.
We shall demonstrate in our study that there are several localization 
patterns and phenomena, when we consider purely only a structural or a combined, structural and diagonal, disorder.

Let's emphasize that we do not impose the symmetry pattern underlying
fragmentation from the very beginning, the symmetries are emerging ones.
Note also that we do not take into account the back reaction of the fermions
on the fluctuating RRG ensemble. The back reaction has been taken into account 
in~\cite{gorsky2023flow}, when the corresponding matrix model has been found which, however, can be treated analytically only at large $N$, corresponding to the
planar approximation of RRG. 

In what follows, we use the following notations. 
The inverse participation ratio $IPR_q$ for the $\psi_i$ eigenstate:
\begin{equation}\label{eq:IPR(E)}
 IPR_{qi}=\sum_{n}^N |\psi_i(n)|^{2q}
\end{equation}
and its average over eigenstates:
\begin{equation}\label{eq:<IPR>}
 IPR_q=\frac{1}{N} \sum_i IPR_{qi}
\end{equation}
\begin{equation}\label{eq:Dq}
 IPR_{q}\sim N^{-\tau_q}, \quad \tau_q=D_q(q-1),
\end{equation}
where $D_q$ are fractal dimensions. For localized states $D_q=0$ and for delocalized states $D_q=1$.

Note that for $q>2$ the presence of finite number of localized states will make the IPR to show non-ergodic properties.
Indeed, if one assumes $N^0$ states to be localized, $IPR_{qi_0}\sim N^0$ and all the rest states to be ergodic, the mean $IPR_q$, with $q>2$ will be dominated by that localized states:
\begin{gather}\label{eq:IPR_q_loc}
IPR_q \propto \frac{1}{N}\lrb{N^0\cdot N^0 + N\cdot N^{-(q-1)}} \sim N^{-1} \lra D_q = \frac{1}{q-1} \ .
\end{gather}
We will use this later as a probe of few localized states in the spectrum.

\subsection{Phase diagram of the model}\label{SubSec:phase_diag}

Now, let's describe the phase structure of the model in $(\mu_k,d)$ parameter plane.
We combine the evident limits with the numerical study of DOS at the
different parameter regimes and different values of $k$.

First, we consider evident limits, having in mind the model~\eqref{eq:eigenproblem}, with the only non-zero chemical potential $\mu_k>0$ at a certain $k$~\eqref{eq:rrg_mu_weights}. 
A pure RRG, with all zero $\mu_k=0$, is a connected graph, with the distribution of cycle lengths $c$ being more or less exponential, 
$P_0(c) \sim (d-1)^c$, which reaches its maximum around the doubled RRG diameter~\cite{bollobas1982diameter}
\begin{gather}\label{eq:diam}
D_{RRG} \simeq \frac{2\ln N}{\ln (d-1)} \ , \quad d>2 \ .
\end{gather}
Then its matrix structure, Fig.~\ref{fig:fig1_phase_diag}(a), contains no clusters, and the corresponding DOS is continuous, Fig.~\ref{fig:fig1_phase_diag}(f), with the only outlier at $\lambda=d$, provided there is the only connected cluster.

\begin{figure}[h!]
 \centering
 \includegraphics[width=0.75\textwidth]{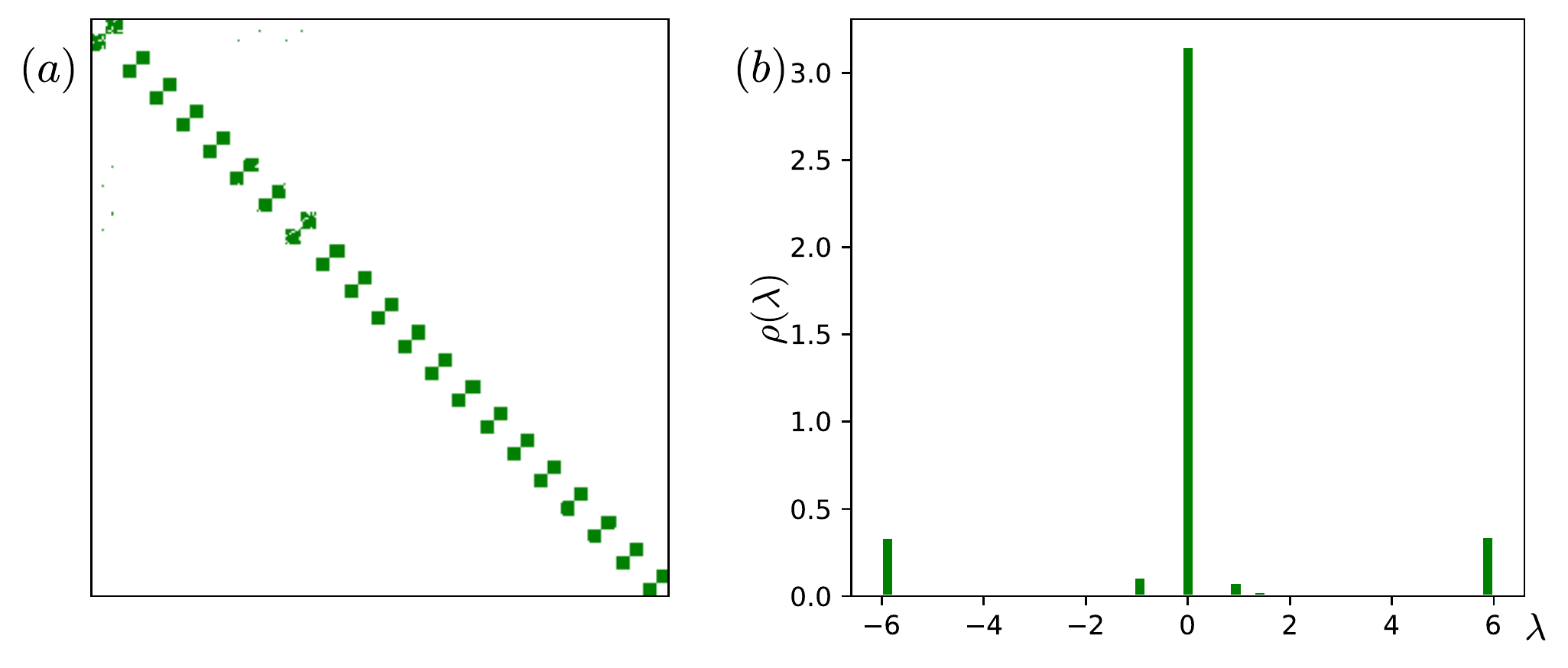}
 \caption{(a) the adjacency matrix, (b) density of states of the graph with 4-cycles chemical potential in the ideally clustered phase.}
 \label{fig:bipartite}
\end{figure}

In the opposite limit of very large $\mu_k\to\infty$, the situation is drastically different as in such a case RRG should split into the disjoint set of identical clusters of size $N_k$ (up to few blocks, given by the remainder of $N$ modulo $N_k$), Fig.~\ref{fig:fig1_phase_diag}(b) for $k=3$ and Fig.~\ref{fig:bipartite} for $k=4$, each of those, being a $d$-regular graph, maximizes the number of $k$-cycles. Here and further, we will refer to such a cluster as an ideal one~\footnote{Note that such ideal clusters exist not for all pairs $(d,k)$.}.
In this case the DOS is highly degenerate, discrete, and it is just given by the spectrum of any of those ideal clusters, Fig.~\ref{fig:fig1_phase_diag}(h), with the degeneracy, proportional to the number of clusters.

Having so different limiting cases, we expect to see a certain kind of a clusterization phase transition between them, which can be unveiled via the DOS profile (if one compares it with the discrete DOS of the ideal-clustered phase), Fig.~\ref{fig:fig1_phase_diag}(d).
The interacting clustered phase has an additional structure in its DOS: in addition to the discrete ideal-cluster DOS, several bands get formed and their number depends on the value of $k$.
The scarred unclustered phase is somehow similar in terms of DOS:
it is characterized by the discrete peaks on top of the continuous DOS of a pure unclustered phase, but now these peaks correspond to graph analogues of scars, given by the so-called topologically equivalent node (TEN) sets. The later do not form a separate cluster, but have several eigenstates, compactly localized on TENs.

Surprisingly from numerics we see a much more rich phase diagram, which shows not $2$, but $4$ distinct phases: in addition to the unclustered phase of pure RRG and the ideally clusterized ones, the interacting-clustered, Fig.~\ref{fig:fig1_phase_diag}(c,i), and the scarred, Fig.~\ref{fig:fig1_phase_diag}(g), phases emerge.

In order to distinct all the $4$ phases, we consider two measures:
\begin{itemize}
 \item One is clusterization measure, namely, the Hellinger distance between the density of states $\rho(\lambda)$ and its ideal-cluster counterpart $\rho_0(\lambda)$
 \begin{equation}
 H_2(\rho,\rho_0) = 1 - \int \sqrt{\rho(\lambda) \rho_0(\lambda)}d\lambda \ ,
 \end{equation}
 which is almost zero in the ideally-clustered phase, jumps to a certain large value in the unclustered phases, and has a certain (rather small) non-zero value in the interacting clustered phase.

 \item Another measure is the fractal dimension, Eq.~\eqref{eq:Dq}, of the high-order $IPR_{q>2}$, Eq.~\eqref{eq:IPR(E)}, averaged over the spectrum, Eq.~\eqref{eq:<IPR>}, which is sensitive to the emergence of localized (scar-like) states in the spectrum, given by the topologically-equivalent node (TEN) sets. This measure distinguishes the unclustered pure and scarred phases, Fig.~\ref{fig:fig1_phase_diag}(e).
\end{itemize}

Analysing the numerical data for $3$-, $4$-, $5$-, and $6$-cycles 
(see Fig.~\ref{fig:fig1_phase_diag} and Figs.~\ref{fig:pd_4cyc}~--~\ref{fig:pd_6cyc} in the Appendix~\ref{App:4-6_cycles}) with help of the above measures, we confirm the universality of the above phase diagram for different $k$-lengths of the cycles and the RRG system sizes $N$. For even $k$ the cluster structure has an additional bipartite property.

$d$-dependence of the clusterization transition $\mu_c$ is always monotonically decaying, while the transition between unclustered pure and TEN-scarred phases, $\mu_{TEN}$, is only slightly growing with $d$ and, thus, crosses $\mu_c$ at a certain $d=d_*$. This $d_*$ is quite close to the one of the ideal-to-interacting clustered transition $d=d_m$ and we conjecture that they should coincide for large enough $N$.
Both transition lines go up with increasing $N$.


In order to estimate the critical lines $\mu_c(N,k,d)$ and $\mu_{TEN}(N,k,d)$, we focus on the limiting cases of pure RRGs and ideal clusters. 
As we have mentioned above, in pure RRG the distribution of cycle lengths $c$ is more or less exponential, and, thus,
the number of cycles at $c=k$ is small for $k\ll D_{RRG}$, $k\gtrsim D_{RRG}$ and most of the cycles are of the length of the doubled diameter $k\simeq D_{RRG}$, Eq.~\eqref{eq:diam}.
This means that the pure RRG looks like a $(d-1)$-degree tree, but with $d$ branches going from a root. As we will see in the next subsection, the structure of the ideal graph for $k$-cycles is very similar.


Another important point, learned from the numerical simulation, is that the peaks in DOS in the scarred unclustered phase are followed by the larger values of $IPR_{q>2}$, signifying the emergence of certain localized eigenstates. They get identified with TENs 
and their positions correspond to the spectrum of the emerging clusters. That is why we 
are tempted to say that TENs are the precursors and nuclei of the clusters.

\subsection{Structure of ideal cluster of $k$-cycles}\label{SubSec:ideal_cluster}
The ideal cluster of $k$-cycles is the cluster, where all smallest (principle) cycles are of the length $k$. As we will see below, not all the parameter pairs $(k,d)$ allow such graphs to exist.

Before we consider a generic case of $k$, let's remind what we know about the structure of ideal clusters for $k=3$ and $k=4$ at fixed $d$:
\begin{itemize}
 \item For $k=3$ the ideal cluster is a complete graph of $d+1$ vertices, see Fig.~\ref{fig:fig1_phase_diag}(b).
 \item For $k=4$ the ideal cluster is a complete bipartite graph of $2$d vertices, see Fig.~\ref{fig:bipartite}.
\end{itemize}
Another simple limiting case for any $k$ is $d=2$: in this situation the ideal cluster is just given by a single $k$-cycle.

For any $k$-cycle ideal cluster we should consider separately odd $k=2n+1$ and even $k=2n+2$ cases.
Following the fact that all the principle (smallest) cycles are of the length $k$ and the vertex degree is $d$, in both cases one should start with the tree of $n$ generations, with $d$ branches of the first generation and the branching number $(d-1)$ for the rest $2$nd, $\ldots$, $n$th generations. Up to now, this construction contains no loops, but the shortest paths between the root and the leaves are of the length 
\begin{equation}
 n = \left[(k-1)/2\right] \ ,
\end{equation}
i.e., the distance between $2$ leaves from different $1$st-generation branches is $2n$.
The difference between even and odd $k$ now is in the connection of leaves. 
\begin{itemize}
 \item For $k=2n+1$ in the ideal cluster one should connect the leaves from different $1$st-generation branches with each other in order to make the links between $n$-paths. 
 There is a way to do it.
 The size of such cluster will be given by the number of vertices in the above tree, i.e.,
\begin{gather}
\label{eq:N_cluster_odd}
N_{k=2n+1} = 1 + d\cdot \sum_{l=0}^{n-1}(d-1)^l = 1 + \frac{d}{d-2}[(d-1)^n-1] \ .
\end{gather}
In this case, the number of $k$-cycles in each cluster can be counted using the above construction.
Indeed, for each of $N_{k}$ vertices, chosen as a root, one should choose a path from the root to one of $d(d-1)^{n-1}$ leaves of $n$th generation plus one of $d-1$ links from this leaf to another one. Then the shortest way back to the root on the tree is certain, as the tree has no loops.
In such a way, we double count each cycle with a fixed root, because both of two tree branches have been counted separately.
In addition, we counted $k$ times each cycle by choosing the root on it.
As a result,
\begin{gather}
\label{eq:N_k-cyc_odd}
N_{cyc,k=2n+1} = N_{k} \frac{d (d-1)^{n}}{2k} \ .
\end{gather}
This formula still has some issues, e.g., for $k=5$ and $d=5m-1$ as the number of cycles in the cluster is not an integer, meaning that there is no ideal cluster for such parameters and some of the principle cycles are of the length, smaller than $k$.

Thus, in the entire $d$-regular graph of $N$ vertices the maximal number of cycles is given by $N N_{cyc,k}/N_{k}$.

 \item For $k=2n+2$ in the ideal cluster one should add another layer of vertices and connect them to the leaves. Counting the number of outgoing edges from the $N_{leaves} = d(d-1)^{n-1}$ leaves as $N_{edges}=(d-1)\cdot N_{leaves} = d(d-1)^n$ and using the regularity of the graph, one will immediately find the number of added vertices as
 \begin{gather}
 N_{add} = N_{edges}/d = (d-1)^n \ ,
 \end{gather}
 leading to the following number of vertices in the even cluster
 \begin{gather}\label{eq:N_cluster_even}
 N_{k=2n+2} = N_{k=2n+1}+N_{add} = \frac{2}{d-2}[(d-1)^{n+1}-1] \ .
 \end{gather}
 In terms of the number of $k$-cycles, this $k$-even case shows similar results with respect to the odd one. Indeed, by choosing the one of $d(d-1)^{n-1}$ leaves of the $n$th generation, one can choose the connection to one of $d-1$ vertices of the additional layer and from it back to one of $d-1$ $n$th-generation leaves, connected to it. As a result, this gives
\begin{gather}
\label{eq:N_k-cyc_even}
N_{cyc,k=2n+2} = N_{k} \frac{d (d-1)^{n+1}}{2k} \ ,
\end{gather} 
which can be generally written for even and odd $k$ as
\begin{gather}
\label{eq:N_k-cyc}
N_{cyc,k} = N_{k} \frac{d (d-1)^{[k/2]]}}{2k} \ ,
\end{gather} 
Like in the $k$-odd case, not all pairs of $(k,d)$ give the integer result for~\eqref{eq:N_k-cyc_even}, e.g., for $n=5$ and $d=5m\pm2$ there is no ideal $10$-cluster: in this case the cluster, maximizing the number of $k$-cycles, should contain some (principle) cycles of the length, smaller than $k$. 
\end{itemize}

Now, knowing the structure of the ideal cluster, it is straightforward to find DOS of the RRG in the ideal-clustered phase.
Indeed, for $k=3$ the $(d+1)\times(d+1)$ adjacency matrix is given by
\begin{gather}
A_3 = \left|\vec{1}\right>\left<\vec{1}\right|-I \ , 
\end{gather}
where $\ket{\vec{1}}\bra{\vec{1}}$ is the rank-one matrix, a projector to a vector of ones $\ket{\vec{1}}$. 
Thus, the eigenvalues of $A_3$ are given by $\lambda_N = d$, corresponding to the eigenvector, equal to the above vector of ones $\ket{\vec{1}}$, and $d$-degenerate eigenvalue $\lambda=-1$, given by the set of eigenvectors, orthogonal to $\ket{\vec{1}}$.

For $k=4$, the ideal cluster is a complete bipartite graph
\begin{gather}
A_4 = \sigma_x\otimes \left|\vec{1}\right>\left<\vec{1}\right| \ , 
\end{gather}
where $\sigma_x$ is a Pauli $2\times2$ matrix, acting in the space of partitions, while all the nodes of different partitions are completely connected to each other.
This matrix has symmetric and antisymmetric solutions $\ket{\pm}\sim (1; \pm 1)^T\otimes\ket{\vec{1}}$, corresponding to $\lambda_\pm = \pm d$, and the rest $2(d-2)$ eigenstates form a complete basis in the space orthogonal to them, with $2(d-2)$-degenerate eigenvalue $\lambda=0$.

The spectra of other $k$-ideal clusters can be also calculated straightforwardly, at least numerically for not very large $N_k$.

\subsection{Estimate of the clusterization transition}\label{SubSec:mu_c}
Focusing on large $N$, $k$, and $d$ further we 
neglect both the cases of non-ideal clusters and a possible incommensurability of $N$ with respect to $N_k$, considering everything in the continuous case. Thus, we assume for simplicity we assume that the number of graph vertices $N$ is a integer multiplier of $N_{k}$ and the number of cycles $N_{cyc,k}$, Eq.~\eqref{eq:N_k-cyc} is also integer.

In this subsection, we estimate the critical chemical potential of clusterization from the straightforward combinatorics by comparing a pure RRG at $\mu_k=0$ with the fully clustered one. 
The entropic factor of number of pure RRG graphs with the fixed number of vertices $N$ and the vertex degree $d$ is given by the following formula for $d\geq3$ and $N\gg d$~\cite{Bender1978_RRG_number}
\begin{gather}
\label{eq:N_RRG}
N_{RRG} = \frac{(d N-1)!! e^{-(d^2-1)/4}}{\left(d!\right)^N}
= \frac{(d N)! e^{-(d^2-1)/4}}{2^{d N/2}(d N/2)!\left(d!\right)^N}\sim \left(\frac{e N}{d}\right)^{d N/2} \ .
\end{gather}
In the last equality we have used the Stirling formula.

At large enough chemical potentials of $k$-cycles all the RRG are represented by the set of $l=N/N_{k}$ ideal clusters of size $N_{k}$.
Thus, the number of these clustered RRG can be estimated as
\begin{equation}\label{eq:N_cRRG}
N_{cRRG} = \frac{N!}{M_k^l} \sim \left(\frac{N}{M}\right)^{N}
 \ ,
\end{equation}
where $M_k$ is the number of indistinguishable ideal clusters, that one can made out of $N_k$ vertices.
For the cases of $k=3$ and $k=4$, on which we will focus in our analytical consideration further
\begin{equation}
M_k = \begin{cases}
(d+1)! \ ,
\quad &k=3 \\
(d!)^2 \ ,
 \quad &k=4
\end{cases}
\ ,
\end{equation}
thus, $M$ in Eq.~\eqref{eq:N_cRRG}, found using the Stirling formula, is given by 
\begin{equation}
M = \begin{cases}
(d+1) \cdot [2\pi(d+1)]^{\frac{1}{2(d+1)}} \ ,
\quad & k=3 \\
d \cdot [2\pi d]^{\frac{1}{2d}} \ ,
 \quad &k=4
\end{cases}
\ .
\end{equation}

At the same time, the number of cycles in the graph grows linearly with $N$ as~\eqref{eq:N_k-cyc} as
\begin{gather}
l N_{cyc,k} = N \frac{N_{cyc,k}}{N_k} \equiv N \frac{d (d-1)^{[k/2]}}{2k} \ ,
\end{gather}
leading to the exponential factor $e^{\mu_k l N_{cyc,k}}$.

Comparing the overall free energy factors of two cases one can estimate (from below) the critical chemical potential $\mu_c$
\begin{gather}
N_{RRG} \simeq N_{cRRG} \cdot e^{\mu_c N_{cyc,k} l} 
\lra
\frac{d}{2}\ln \left(\frac{e N}{d}\right) = \ln \left(\frac{N}{M}\right) + \mu_c \frac{N_{cyc,k}}{N_{k}} \ ,
\end{gather}
where the last equality is obtained by taking the logarithm and dividing by $N$.
As a result, this critical chemical potential value takes the form
\begin{gather}\label{eq:mu_c(N,k,d)}
\mu_c(N,k,d) \simeq k\frac{(d-2)\ln N+2\ln M - d \ln (d/e)}{d(d-1)^{[k/2]}} \ ,
\end{gather}
which at large enough $N$ grows logarithmically with $N$.

Note that the above estimate should be a lower bound for the clustering transition as we compare the ideal-clustered phase only with the pure RRG at $\mu_k=0$, neglecting other possible contributions at finite $\mu_k>0$.
Nevertheless, this estimate give a reasonable quantitative agreement with the numerics, see the solid red line in Fig.~\ref{fig:fig1_phase_diag}(d).

\section{Quantum scars in the spectrum and topological equivalent nodes (TEN)}\label{Sec:TENs}
In this Section we demonstrate that there are interesting quantum-scar 
localized states in the spectrum of perturbed RRG ensemble and describe their properties.
They have some underlying symmetry origin, but situation differs considerably from the approach of~\cite{pakrouski2020many}, when 
the scars also emerge from the representation group of a certain symmetry.
In that case, the symmetry is built-in into the Hamiltonian, while in our case it is a sort of induced upon perturbation. 

In Sec.~\ref{SubSec:num_scars=TENs} we define our TEN-scar states and provide a numerical evidence of them. 
In the next subsections we consider analytically the origin of TENs and their spectral locations.
We start with the unclustered phase and simple multipole TENs at $\lambda=-1$ for $k=3$ ($\lambda=0$ for $k=4$), Sec.~\ref{SubSec:TEN_E=-1}, followed by the estimate of the TEN-scar transition in the unclustered phase, Sec.~\ref{SubSec:mu_TEN}.
And then switch to more complicated structure of interacting TEN clusters, leading to
the emergence an additional TEN band around $\lambda=-1$ for $k=3$, Sec.~\ref{SubSec:TEN_band}, and peaks at $\lambda=0,\pm\sqrt{n}$ for $k=4$, $6$, Sec.~\ref{SubSec:TEN_even_sqrt_n}.

\subsection{On the definition and identification of TENs}\label{SubSec:num_scars=TENs}
The scar states are roughly speaking localized at the nodes that are similarly connected to
the environment - the notion TEN follows from this property. 
The number of nodes the scar is localized at can be different and 
we will find dipole and multipole scars at several nodes. The definition 
for the multipole scars is relatively simple, however, there are 
more complicated "scar subgraph", when it is difficult to formulate universally. We shall present below the detailed analysis of the 
simple scar multipoles. The examples of more complicated "scar subgraph", when only the necessary condition for TEN can be rigorously formulated, is given in Appendix~\ref{App:TEN_complexes}.

Simple multipole TENs are defined as the set of nodes, for which all the connections to the nodes, outside the set, are common (and identical) for all set nodes. Note that in such a case, the internal connections within the node set can be arbitrary.


\begin{figure}[h!]
\centering
\includegraphics[width=.95\textwidth]{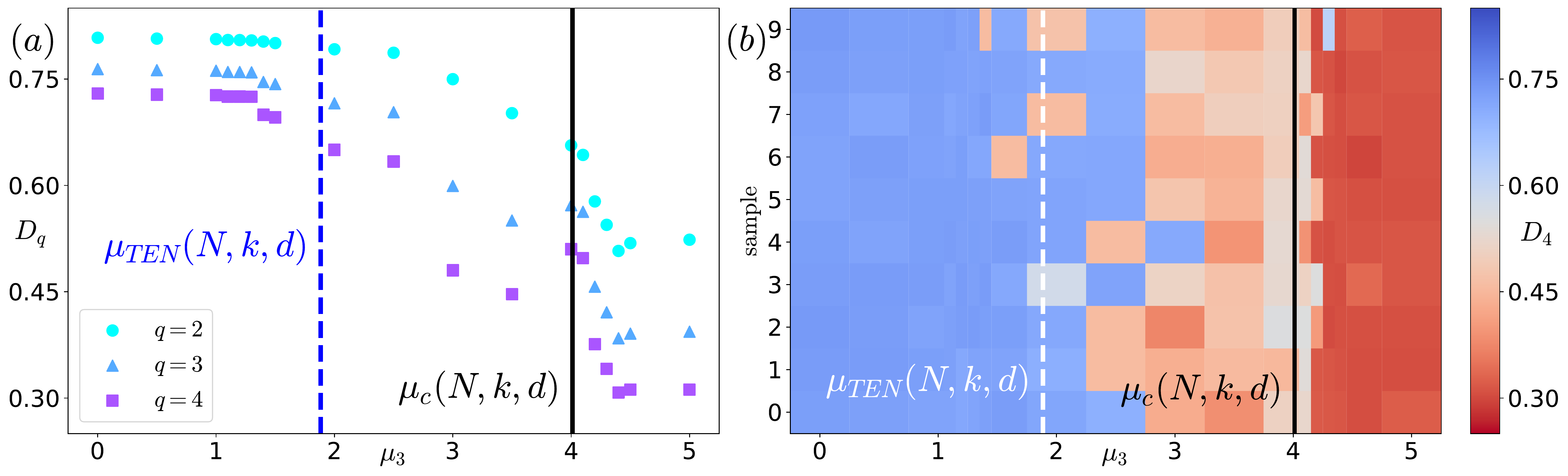}
\caption{Fractal dimension $D_q$ ($IPR_q\sim N^{-(q-1) D_q}$) vs the chemical potential $\mu_k$ for $N=256$, $k=3$, and $d=3$, plotted (a)~for different $q$ (color symbols) 
and (b)~for different realizations of the sample the jump in $D_4$ appears at slightly different $\mu_k$. The vertical lines show the transitions of first scar emergence, $\mu_{TEN}$, and the clusterization, $\mu_c$.}
\label{fig:RRG_IPR_scar_jump}
\end{figure}


Let us explain how scars can be identified numerically in a simple manner.
Take a closer look at the $\mu$-dependence of the higher-order $IPR_{q>2}$, see Fig.~\ref{fig:fig1_phase_diag}(e). In figure~\ref{fig:RRG_IPR_scar_jump}, we show a cross-section of the above diagram at a certain $d$: $IPR_{q>2}$ has a jump discontinuity at an emergence of the first scar. That jump appears due to a special (localized) state which dominates the average $IPR_q$ value, see Eq.~\eqref{eq:IPR_q_loc}, and the amplitude of it drastically increases with $q$.

\begin{figure}[h!]
\centering
\includegraphics[width=1\textwidth]{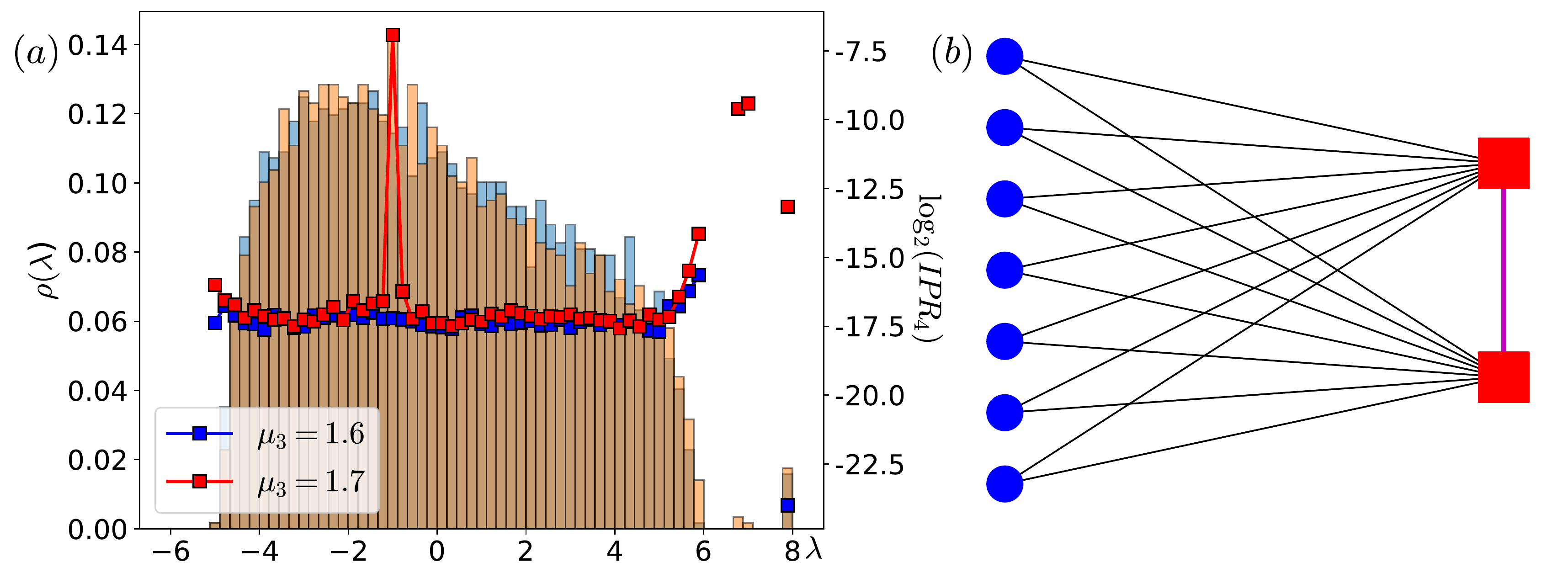}
\caption{(a) Evidence of TEN emergence in the density of states (color histograms) and energy-resolved $IPR_4$ (symbols) for $k=3$-cycles. State before (after) jump discontinuity vs $\mu_k$ is marked by blue (orange) color. The amplitude of the local maximum at $\lambda=-1$ of both DOS and $IPR_4$ after the jump drastically increases. (b)~Dipole TEN nodes (red squares) with the nearest neighbors (blue circles), that form a DOS peak at $\lambda=-1$. Edge, connecting TEN nodes is highlighted in magenta.}
\label{fig:3_cycle_eigIPR}
\end{figure}

It is that state which is located at the center of the mid-spectrum perturbative zone at the eigenvalue $\lambda=-1$ for $k=3$, where the DOS peak appears at larger $\mu$, Fig.~\ref{fig:3_cycle_eigIPR}. The eigenvector of that state has only two nonzero elements, equal by absolute value and opposite in sign.
All neighbors of the nonzero vertices are common, that leads to indistinguishability of the nodes for the external network and such configuration bears the name of the topologically equivalent nodes (TEN). Such topological property gives the localization and the nodes form an effective dipole. 

Creation of the TEN is natural for a network, with large enough triangle chemical potential, $\mu_3$, in a clustered phase. 
Indeed, a pair of connected nodes and each of their common neighbors form a triangle. Because of clustering, the number of possible common neighbors is limited since cluster size is approximately $\sim (d+1)$. 
However, as we see from Fig.~\ref{fig:fig1_phase_diag} the first TENs emerge even before the clusterization transition and work as nuclei and precursors of these clusters.

The larger the chemical potential $\mu_k$ is taken, the more the number of dipole state increases. In addition to the above dipolar states, states with the larger number of nodes in TEN-cluster form. Among them, there are simple multipoles and more complex states. 


\begin{figure}[h!]
\centering
\includegraphics[width=\textwidth]{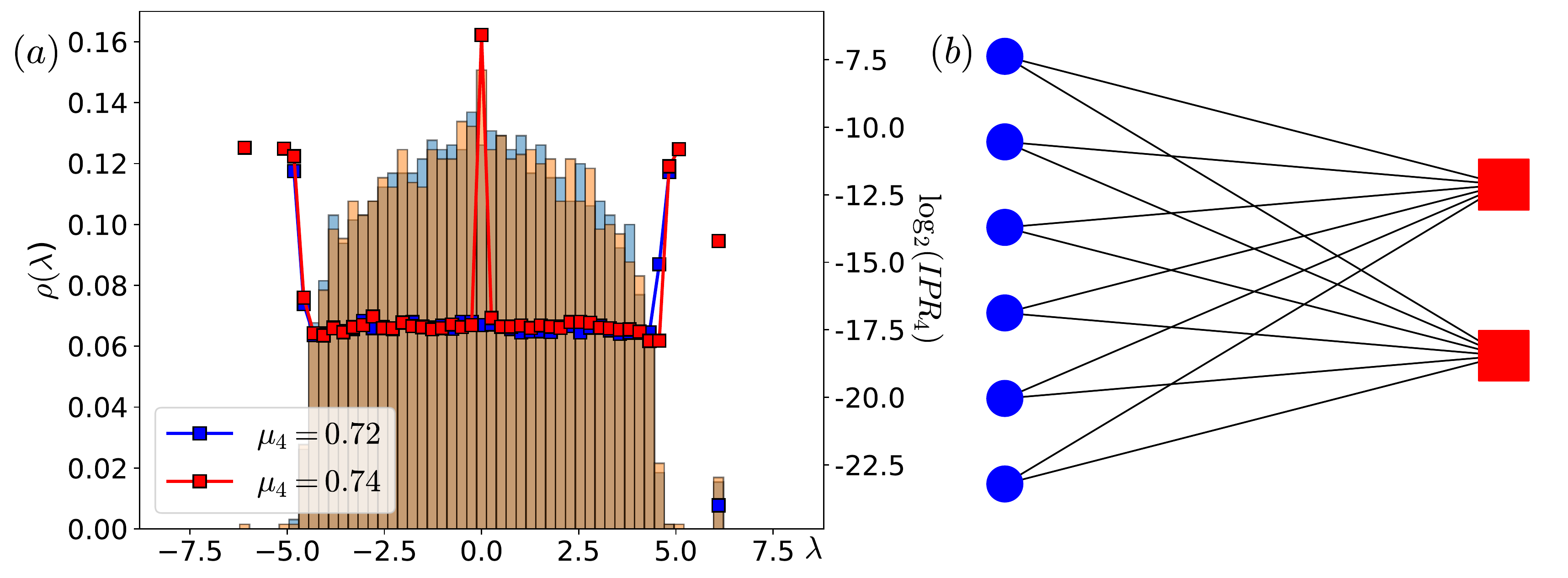}
\caption{(a)~Evidence of TEN emergence in the density of states (color histograms) and energy-resolved $IPR_4$ (symbols) for $k=4$-cycles at $\lambda=0$. The notations are the same as in Fig.~\ref{fig:3_cycle_eigIPR}. (b)~Dipole TEN nodes (red squares) with the nearest neighbors (blue circles), that form a DOS peak at $\lambda=0$.}
\label{fig:four_0}
\end{figure}

Similar TENs exist also in $k=4$-cycles. Clustered phase for a graph with $4$-cycle chemical potential $\mu_4$ tends to be bipartite and DOS becomes symmetric with respect to $\lambda=0$. 
The first emerging TENs, appearing before the clusterization, at $\lambda=0$ do not obey this bipartite symmetry, Fig.~\ref{fig:four_0}, but in terms of the spectrum provide the basis for its appearance at higher $\mu_4$. Indeed, being a dipole (multipole) in one partition of a graph which is about to become bipartite, these TEN nodes are disconnected from each other, having instead all $d$ common neighbors in the other partition. Thus, the TEN eigenstate, localized compactly on the TEN nodes, lives only in the one partition.

\begin{figure}[h!]
\centering
\includegraphics[width=\textwidth]{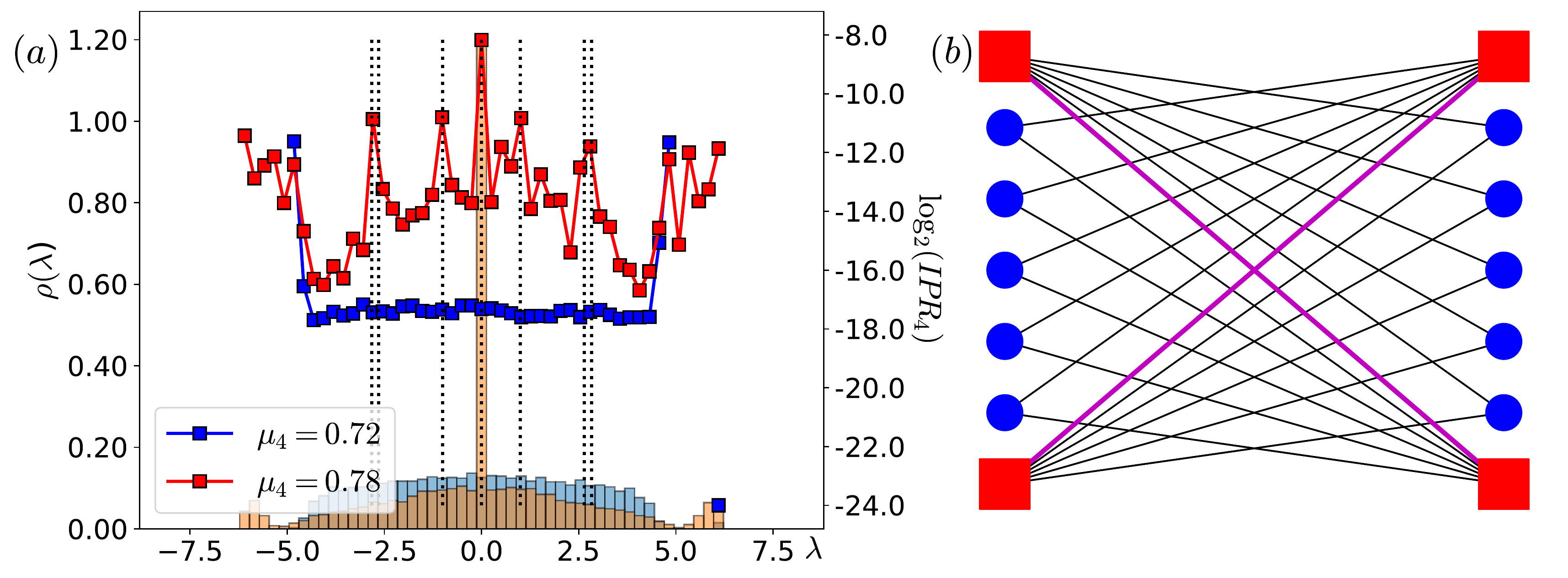}
\caption{(a)~Evidence of additional TEN emergence at $\lambda=\pm 1$, $\pm\sqrt{7}$, $\pm\sqrt{8}$ (vertical dashed lines) in the density of states (color histograms) and energy-resolved $IPR_4$ (symbols) for $k=4$-cycles. The notations are the same as in Fig.~\ref{fig:3_cycle_eigIPR}. (b)~Two coupled TEN sets (red squares), with the nearest neighbors (blue circles), that form a DOS peak at $\lambda=\pm1$. Edges, connecting TEN sets are highlighted in magenta.}
\label{fig:four_pm1}
\end{figure}

Next, at larger $\mu_4$ and $d$ values in the interacting clustered phase, other TEN complexes emerge symmetrically in the spectrum, see them at $\lambda=\pm1$ in Fig.~\ref{fig:four_pm1}. The most significant peaks in DOS appear to be at $\lambda=\pm \sqrt{n}$ for even cycles.
These peaks correspond to the interacting simple TEN dipoles (multipoles), with a certain inter-TEN edge structure, see Fig.~\ref{fig:four_pm1}(b).


Above, we have presented some examples of a successful scar hunting in
RRG, perturbed by $\mu_3$ and $\mu_4$. The same procedure allows to identify scars for higher cycles as well, see Appendix~\ref{App:4-6_cycles}. The case of TENs in the interacting clustered phase has been considered above, however the inspection of peaks in the non-clusterized phase at small chemical potentials also recovers their scar nature.

\subsection{Origin of first TEN formation in $3$- and $4$-cycles}\label{SubSec:TEN_E=-1}
Focusing on the $k=3$- and $k=4$-cycle cases, we remember that from the numerics there was the difference between that cases. Indeed, the first $3$-cycle TENs emerge at $\lambda=-1$, while for the $4$-cycle case it, first, happens at $\lambda=0$. What is the difference between these cases?

Let's consider the simplest TEN multipole of $n$ nodes, where all the nodes, $l=1,\ldots,n$, in the TEN set are topologically equivalent not only to the external nodes, but also between each other.
As each pair of TEN nodes is connected (or not) to all the rest of the nodes in the network identically, we have only two options within the set:
\begin{itemize}
 \item Either all of TEN nodes are connected to each other, 
 \item or all of them are disconnected. 
\end{itemize}

In the former case we get the following equations
\begin{gather}\label{eq:TEN_k=3}
\lambda \psi_l = \sum_{l'=1}^{n} \psi_{l'} - \psi_l +S \ , \quad 1\leq l\leq n \ ,
\end{gather}
where $\psi_l$ are the wave-function components on the TEN set, $\phi_i$ are the ones on the common neighboring nodes, and $S = \sum_{i=1}^{d-n} \phi_i$. 
Summing over $l$ the above equations, we immediately obtain
\begin{gather}
S = (\lambda-n+1) b \ ,
\end{gather}
where 
\begin{gather}\label{eq:b=sum_bl}
b = \frac{1}{n}\sum_{l=1}^{n} \psi_l \ . 
\end{gather}
Taking into account the fact that in the equations for the neighboring nodes $\phi_i$ the TEN-set ones are always coming as $n\cdot b$, one can straightforwardly get that for all the TEN modes with $b=0$ the coupling to the nodes outside the cluster is zero, $S = 0$. 

Taking this condition $S=b=0$ into account, one immediately obtains the solution at $\lambda=-1$ from~\eqref{eq:TEN_k=3}. Therefore we have $n-1$ TEN modes $(\psi_1,\ldots,\psi_{n})$ at $\lambda=-1$, which are orthogonal to the vector of unities:
\begin{gather}\label{eq:sum=0}
\brakett{\vec{1}}{\vec{b}}=\sum_{l=1}^{n} \psi_l = 0 \ . 
\end{gather}
Here we used the fact that for TEN the coefficients $\psi_{l>n}\equiv 0$. 


In the latter opposite case of all disconnected states, they form the degenerate states at $\lambda=0$ since
\begin{gather}
\lambda \psi_l = S \ , \quad 1\leq l\leq n \ ,
\end{gather}
where $S=\sum_{i=1}^{d} \phi_i$. Indeed, the summation over all $l$ gives $S=\lambda b$ and 
the condition~\eqref{eq:sum=0} emerges in this case as well, leading to $n-1$ TEN modes at $\lambda=0$.

It is clear that in the case of positive chemical potential $\mu_3$ of $3$-cycles the formation of links between TEN-set nodes is preferable since it increases the number of triangles. That is why the states of the first type at $ \lambda=-1$ get emerged as we have observed in the simulations. 

Similarly in the case of $k=4$, the preferred state is to have TEN nodes, completely disconnected from each other and connected only to the common neighbors, as this increases the number of $4$-cycles. Such a reasoning explains the emergence of $\lambda=0$ degenerate states in the case of $k=4$. 

\subsection{Estimate of the TEN-scarred transition}\label{SubSec:mu_TEN}
Now, knowing the origin of the first TEN states, we are ready to estimate the chemical potential $\mu_{TEN}$, when they first emerge, see Fig.~\ref{fig:fig1_phase_diag}(e).

Let's, first, focus on the case of $k=3$-cycles, where the first TEN appears, as soon as one of the neighbors $j$ of a certain node $i$ connects to all the same $d-1$ other neighbors of $i$.
In such a way, the number of $k=3$-cycles increases by $d-1$, while the number of such RRG with the only TEN can be estimated in the similar way as in~\eqref{eq:N_RRG},~\cite{Bender1978_RRG_number}.
Up to an exponential prefactor $e^{-(d^2-1)/4}$, expression~\eqref{eq:N_RRG} can be calculated from a straightforward construction (see, e.g., Sec. 9.1 of~\cite{Bollobas2001random_graphs}): for the $d$-regular graph, we consider
a $N\times d$ table, containing $d$ times every number from $1$ to $N$.
Then (up to the self-loops and repeated edges, taken into account by the exponential prefactor $e^{-(d^2-1)/4}$), the number of RRG realizations is given by the number of node pairs (i.e., edges), chosen from that table $(d N-1)!!$, factorized by the permutation of within each of $N$ sets of $d$ equal numbers, $(d!)^N$.

In the case of the first TEN for $k=3$, we should slightly modify the above procedure.
First, like in the pure RRG case, let's choose all $d$ neighbours for a certain node $i$, which gives us $(d N-1)!!/(d N-2d-1)!!$ variants.
Then we choose one of its $d$ neighbors $j$ and $(d-1)!$ possibilities to connect it to the {\it same} $d-1$ neighbors of $i$. The latter gives the factor of $d!$.
Then, the rest $(d N-4d)$ edges can be taken randomly, like in the pure RRG case, giving $(d N - 4d-1)!!/(d!)^N$. The only thing which one should take into account is that the neighbors $i$ and $j$ can be, in general, taken out of $N(N-1)/2$ variants, but not only being the first and the second.
Summarizing all these estimates, one should have the number of RRG with a single TEN in $k=3$, given by
\begin{multline}\label{eq:N_TEN}
N_{RRG,TEN} = \frac{(d N-1)!!(d N-4d-1)!! N(N-1)}{(d N-2d-1)!! (d!)^{N-1} 2} =\\= N_{RRG} \frac{d! N(N-1)}{2(d N-2d-1)(d N - 2d - 3)\cdot\ldots\cdot (d N - 4d+1)} \simeq N_{RRG} \frac{\sqrt{\pi d/2}}{N^{d-2} e^d} \ .
\end{multline}
The latter approximate equality assumes $N\gg 1$ and uses the Stirling formula.
Note that, like Eq.~\eqref{eq:N_RRG}, here we know this number up to an exponential factor of some polynomial of $d$, due to the self-loops and repeated edges.

As a result, comparing
\begin{equation}
N_{RRG} = N_{RRG,TEN} e^{\mu_{TEN} (d-1)}
\end{equation}
one obtains
\begin{equation}\label{eq:mu_TEN}
\mu_{TEN,3} = \frac{1}{d-1}\left[(d-2)\ln N + d - \ln \sqrt{\pi d/2} - P(d) \right] \ ,
\end{equation}
where the unknown polynomial $P(d)$ appears due to the unknown exponential in $d$ prefactor from Eq.~\eqref{eq:N_TEN}.
In order to estimate it we fit it as $P(d)\simeq 4(d-2)$ from the data at one system size $N=512$ and show the result for the other one, $N=256$ in Fig.~\ref{fig:fig1_phase_diag}(d,e).

The case of $k=4$-cycles is pretty similar, but now, according to Sec.~\ref{SubSec:TEN_E=-1}, the first TEN set is given by $2$ disconnected nodes with all $d$ common neighbors.
Surprisingly, following the same above procedure one obtains the same expression for $N_{RRG,TEN}$ as in Eq.~\eqref{eq:N_TEN}, while the number of created $4$-cycles is given by $d(d-1)/2$, leading to
\begin{equation}\label{eq:mu_TEN4}
\mu_{TEN,4} = \frac{2}{d(d-1)}\left[(d-2)\ln N + d - \ln \sqrt{\pi d/2} - P'(d) \right] \ ,
\end{equation}
where $P'(d)$ may be in principle different from the above $P(d)$.

Here one should notice that both estimates~\eqref{eq:mu_TEN} and~\eqref{eq:mu_TEN4} are valid until $\mu_{TEN,k}<\mu_c(N,k,d)$, Eq.~\eqref{eq:mu_c(N,k,d)}.
Indeed, as soon as we assume the emergence of the first TENs to be on the background of a pure RRG, the latter should be unclustered.
Thus, at $\mu_{TEN,k} = \mu_c(N,k,d)$ something dramatic should happen.
As soon as the clusterization happens before the emergence of the first ``simple'' TENs, such a process should involve some non-ideal or interacting clusters. This is what 
we, indeed, see from our phase diagram in Fig.~\ref{fig:fig1_phase_diag}. 

In addition, surprisingly, in both cases of $k=3$ and $k=4$
the intersection of $\mu_{TEN,k}$ and $\mu_c(N,k,d)$ gives
\begin{gather}
d = 3 + O\left(\frac{1}{\ln N}\right) \ ,
\end{gather}
thus, only at $d=3$ in the thermodynamic limit one can see the effects of TENs in the unclustered phase.

In the case of interacting clustering phase, we still see the peaks at the energy $\lambda=-1$ ($\lambda=0$) for $k=3$($k=4$)-cycles, but they are supplemented by some non-trivial bands of states. For $k=3$ it is some kind of the triangular-shaped DOS, see Fig.~\ref{fig:3_cycle_eigIPR}, while for $k=4$ and $k=6$, the additional peaks appear at $\lambda=\pm \sqrt{n}$.
In the next two subsections we show the origin of these phenomena.

\subsection{Band formation of TENs 
}\label{SubSec:TEN_band}
If the nodes $1\leq l\leq n$ of the TEN-cluster are topologically equivalent only with respect to the outer nodes $j>n$ and wave functions are localized at nodes $l$, the condition~\eqref{eq:sum=0} gets emerged as well. 
Indeed, for any adjacency matrix $A_{ll'}$, $1\leq l,l'\leq n$ within the TEN set of nodes one has
\begin{gather}\label{eq:psi_l_genTEN}
\lambda \psi_l = \sum_{l'=1}^{n} A_{ll'}\psi_{l'} + S \ , \quad 1\leq l\leq n \ . 
\end{gather}
Note that here 
\begin{gather}\label{eq:S_sum_mi}
S = \sum_{i=1}^{d-d_n} \phi_i
\end{gather} 
is the sum over the same amount of external nodes $\phi_i$, i. e., for the regular graph with the degree $d$, the internal adjacency matrix also represents a regular graph, with a certain degree $d_n$, i. e.,
\begin{gather}\label{eq:sum_A_ll'}
\sum_l A_{ll'} = d_n \ . 
\end{gather}

Let's focus, first, on the RRG case for $A_{ll'}$, i. e. , when Eq.~\eqref{eq:sum_A_ll'} is valid (This does not mean that the entire graph is regular, but just all the TEN-cluster nodes have the same degree $d_n$). 
Summing Eq.~\eqref{eq:psi_l_genTEN} over all $1\leq l\leq n$ and taking into account~\eqref{eq:sum_A_ll'}, one immediately obtains that $S$ again depends only on the sum~\eqref{eq:b=sum_bl}, appearing in all equations for $\phi_i$
\begin{gather}
S = (\lambda-d_n) b \ . 
\end{gather}

Following the same procedure as in Sec.~\ref{SubSec:TEN_E=-1}, one can show that all the $n-1$ modes in the cluster, obeying the condition $b=0$, Eq.~\eqref{eq:sum=0}, are TENs, i. e., they are decoupled from the rest of the nodes, $S=0$. 

The spectrum of these TENs then is given by the spectrum of the $n\times n$ (random) regular graph $A_{ll'}$ with the degree $d_n$, except for the only eigenstate, with $b\ne 0$ and $\lambda = d_n$, which is symmetric with respect to the internal node permutations
$(\psi_1,\ldots, \psi_{n}) \propto (1,\ldots,1)$. 

As a result, such additional TENs with the internal structure of the node set probably form a triangular-shaped DOS in the interacting clusterized phase.

\subsection{Origin of TEN at $\lambda=\pm \sqrt{n}$ for bipartite $4$- and $6$-cycles}\label{SubSec:TEN_even_sqrt_n}
For even cycles, it is numerically observed that the system turns to be bipartite at $\mu>\mu_c(N,k,d)$.
However, in addition to the standard TEN states at $\lambda=0$, in the interacting clustered phase there appear also rather strong peaks at $\lambda=\pm \sqrt{n}$, see Fig.~\ref{fig:pd_6cyc} in Appendix~\ref{App:4-6_cycles}. 

In order to uncover the origin of such state, we consider two sets of nodes $(\psi_1,\ldots,\psi_{n})$ and $(\vphi_1,\ldots,\vphi_{m})$ from different bipartite components, that form TENs.
Without loss of generality here we consider $n\leq m$.

Again we assume that each set is topologically equivalent to all the outer nodes (excluding the bipartite partners). 
Taking into account the bipartite character of the graph and putting a certain adjacency matrix $A_{lj}$ between $\psi_l$ and $c_j$ modes,
one can obtain the following equations
\begin{eqnarray}
\label{eq:b_l_bipartite}
\lambda \psi_l &=& \sum_{j=1}^{m}A_{lj} \vphi_{j} + S_b \ ,\\
\label{eq:c_j_bipartite}
\lambda \vphi_j &=& \sum_{l=1}^{n} A^T_{jl} \psi_{l} + S_c 
\ ,
\end{eqnarray}
where $S_{b,c}$ are the couplings to outer nodes with respect to two above sets and, thus, $A_{lj}$ is regular in both indices, i.e.,
\begin{gather}\label{eq:A_lj_sum}
\sum_l A_{lj} = d_m \ , \quad 
\sum_j A_{lj} = d_n \ , \quad
d_m\cdot m = d_n\cdot n \ .
\end{gather}
Note that, as we have positive chemical potential of even cycles, like in Sec.~\ref{SubSec:TEN_E=-1}, the most probable case is when all $\psi_l$ are coupled to all $\vphi_j$, i.e., $d_m = n$ and $d_n = m$.

Introducing again the variables
\begin{gather}
b = \frac{1}{n}\sum_{l=1}^{n} \psi_l \ , \quad 
c = \frac{1}{m}\sum_{j=1}^{m} \vphi_j 
\end{gather}
and using the same trick of summation of the equations~\eqref{eq:b_l_bipartite},~\eqref{eq:c_j_bipartite} over indices $l$ and $j$, respectively, one can obtain with help of~\eqref{eq:A_lj_sum} the following equations for $S_{b,c}$
\begin{gather}\label{eq:Sb_Sc_via_b_c}
S_b = \lambda b - d_m c \ , \quad
S_c = \lambda c - d_n b \ . \quad
\end{gather}
In order to have TEN, one should have $S_b = S_c = 0$ from the equations for $\psi_l$ and $\vphi_j$ and $b=c=0$ from the equations for nodes, hidden in $S_b$ and $S_c$.

As a result, for any TEN of this bipartite kind, from $b=c=0$ one immediately obtains from~\eqref{eq:Sb_Sc_via_b_c} $S_b = S_c = 0$.

Returning back to the equations~\eqref{eq:b_l_bipartite} and~\eqref{eq:c_j_bipartite}, similarly to the previous section, we should find the (left and right) eigenstates of the internal adjacency matrix $A_{lj}$, consistent with the condition $b=c=0$.

By substituting the expression of $\vphi_j$ from~\eqref{eq:c_j_bipartite} to~\eqref{eq:b_l_bipartite}, one can straightforwardly obtain that the eigenvalues are given by the solution of the following problem
\begin{gather}\label{eq:b_l_bipartite_solution}
\lambda^2 \psi_l = \sum_{l'} B_{ll'} \psi_{l'} \ ,
\end{gather}
where
\begin{gather}
B_{ll'} = \sum_j A_{lj}A^T_{jl'} \ , \quad
B_{ll} = \sum_j A_{lj} = d_n \ , \quad
\sum_l B_{ll'}=d_n\cdot d_m \ .
\end{gather}
In the second equality we used the fact that $A_{lj} = A^T_{jl}$ and can take only two values $0$ and $1$.

For the most frequent case of all-to-all coupling $A_{lj}=1$, $d_n=m$, $d_m=n$, one immediately obtain the $n+m-2$ standard independent TENs of either of two types at $\lambda=0$, considered in Sec.~\ref{SubSec:TEN_E=-1}:
\begin{eqnarray}
b &=& 0 \ , \quad \vphi_j = 0\\
c &=& 0 \ , \quad \psi_l = 0 \ .
\end{eqnarray}

For the simple case of $n=2$ and $d_m=1\leq n$, see the example in Fig.~\ref{fig:four_pm1} for $d_n=d_m=1$, $m=n=2$, one can immediately find from $b=0$ that
$\psi_2 = -\psi_1$ and the equation~\eqref{eq:b_l_bipartite_solution} has the only solution
\begin{gather}
\lambda^2 = B_{11}-B_{12} = 2 B_{11}-\sum_{l} B_{ll'} = d_n\cdot(2-d_m) = d_n \ ,
\end{gather}
with a certain integer $d_n\leq m$.

Larger $n$ will give more opportunities and will form the band around peaks at $\lambda=\pm\sqrt{d_n}$.

\section{Localization properties of non-perturbative band}\label{Sec:semi-Poisson_side_band}

 It has been argued in~\cite{avetisov2020localization} that in the 
 non-perturbative side-band of the Erdos-Renyi model, perturbed by $\mu_k$, with the constraint of fixed initial vertex degrees, the eigenvalues obey almost the Poisson statistics. Based on that, it has been conjectured that these states are localized. In this Section, we investigate this point more carefully in the corresponding interacting clustered phase of the $\mu_k$-perturbed RRG by analysing the level spacing distinction (LS), the fractal dimension, and IPR. It turns out that the spectrum in non-perturbative side-band obeys the {\it semi-Poisson} statistics. 
 LS determine statistic of spacing between two adjacent energy levels
\begin{equation}
 s_i=E_{i+1}-E_i,
\end{equation}
where $E_{i}$ are energy levels after unfolding. 
Energy levels after unfolding
\begin{equation}\label{eq:E_i_unfolded}
 E_i=\mathrm{CDF}(\lambda_i)N
\end{equation}
have been calculated from the raw spectrum $\lambda_i$, $1\leq i\leq N$, using the standard calculation of a cumulative distribution function (CDF), given by a fraction of eigenvalues smaller than $\lambda$, averaged over an ensemble of $n$ graph realizations, $\lambda_i^{(j)}$, $1\leq j\leq n$:
\begin{equation}
\mathrm{CDF}(\lambda)=\frac{\#(\lambda_i^{(j)}<\lambda)}{nN} \ .
\end{equation}

Semi-Poisson is interpolation between the GOE distribution and the Poisson distribution~\cite{Bogomolny1999models}. Generalised semi-Poisson distribution
\begin{equation}
P(s)=C_1(\gamma, \alpha) s^\alpha e^{-C_2(\gamma, \alpha) s^{2-\gamma}}
\end{equation}
where
\begin{equation}\label{eq:gen_semi-Poisson}
C_2(\gamma, \alpha)=\left(\frac{\Gamma\left(\frac{2+\alpha}{2-\gamma}\right)}{\Gamma\left(\frac{1+\alpha}{2-\gamma}\right)}\right)^{2-\gamma} \quad \text{and} \quad C_1(\gamma, \alpha)=\frac{(2-\gamma) C_2^{\frac{1+a}{2-\gamma}}}{\Gamma\left(\frac{1+\alpha}{2-\gamma}\right)}
\end{equation}
\begin{figure}[h!]
 \centering
 \includegraphics[width=1\textwidth]{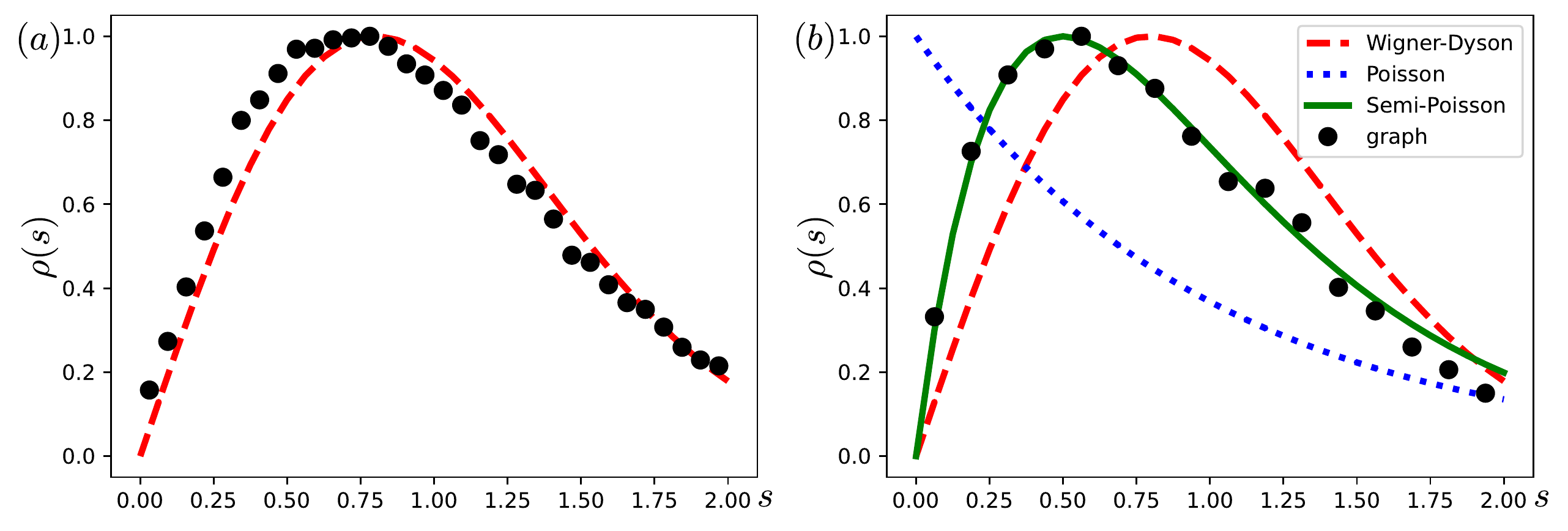}
 \caption{Level spacing distribution for RRG in clustered phase for $k=3$, $d=20$, $N=256$ separately for (a)~the perturbative mid-spectrum band ignoring TEN states and (b)~the non-perturbative side-band. Red dashed, blue dotted, and green solid lines show Wigner-Dyson ($\alpha=1$, $\gamma=0$), Poisson ($\alpha=0$, $\gamma=1$), and semi-Poisson ($\alpha=1$, $\gamma=1$) distributions, Eq.~\eqref{eq:gen_semi-Poisson}, respectively.
Level spacing is calculated for the unfolded spectrum, Eq.~\eqref{eq:E_i_unfolded}, using $500$ random realisations.}
 \label{fig:lsd}
\end{figure}
The Poisson statistic corresponds to $\alpha=0$ and $\gamma=1$, the Wigner-Dyson distribution to $\alpha=1$ and $\gamma=0$. In Fig.~\ref{fig:lsd}, the level statistics in the interacting clustered phase is shown both for the mid-spectral ignoring TEN states and side bands. While the mid-spectral band shows Wigner-Dyson statistics, semi-Poisson one with $\alpha=1$ and $\gamma=1$ appears in the side band. 

Usually the semi-Poisson distribution, with $\alpha=1$ and $\gamma=1$, appears as a result of some constraint~\cite{Bogomolny1999models} or Bethe-ansatz integrability~\cite{Yuzbashyan_NJP2016}: indeed, if $\epsilon_i$, $0\leq i \leq N$ are Poisson distributed random numbers, then $(\epsilon_{i-1}+\epsilon_i)/2$ sample the semi-Poisson distribution.

As soon as the non-perturbative side-band in the clustered phases of the perturbed RRG is composed out of the homogeneous modes on each of the separate clusters, all such modes in the ideal clustered phase are degenerate at the cluster size $\lambda = N_k$, see Eqs.~\eqref{eq:N_cluster_odd} and~\eqref{eq:N_cluster_even}.
With increasing $d$ the RRG undergoes the transition to the interacting clustered phase, where the above clusters are not disjoint anymore: they interact with each other, lifting the degeneracy of the non-perturbative side-band.

Interaction between clusters is given by the edges between nodes of different clusters.
Internally each of non-ideal clusters in this phase is similar to the TEN set and therefore is almost a regular graph with the degree $d-d_0$, which fluctuates from cluster to cluster. Thus, each node of it is connected randomly on average to the $\mean{d_0}$ other clusters.
In the same way, each of the homogeneous cluster modes (that were at $\lambda=N_k$ in the ideal clustered phase) is connected on average to $N_k\cdot \mean{d_0}$ of other cluster modes.
As soon as the number of clusters $N/N_k$ is less compared to $N_k \cdot \mean{d_0}$, i.e., at
\begin{gather}
 \mean{d_0}\geq \frac{N}{N_k^2} \ ,
\end{gather}
all such modes are almost all-to-all coupled, like in the Bethe-ansatz integrable Richardson's model~\cite{Yuzbashyan_NJP2016,Nosov2019mixtures}.
Fluctuations in $d_0$ shift the cluster mode energy with respect to $N_k$ and then play a role of the effective Poisson disorder potential $\epsilon_i$.
Thus, effectively the structure of the non-perturbative side-band, formed by the cluster modes, mimics the one of the Richardson model.
In the latter, from the Bethe-ansatz solution it is known that the eigenvalues are squeezed between the diagonal potential values $\epsilon_{i-1}<\lambda_i<\epsilon_i$, sorted over their values, and the all-to-all coupling pushes $\lambda_i$ away from $\epsilon_j$. In a strong coupling limit, this leads to $\lambda_i \simeq (\epsilon_i+\epsilon_{i-1})/2$, confirming the semi-Poisson nature of the level statistics.

\begin{figure}[h!]
 \centering
 \includegraphics[width=0.7\textwidth]{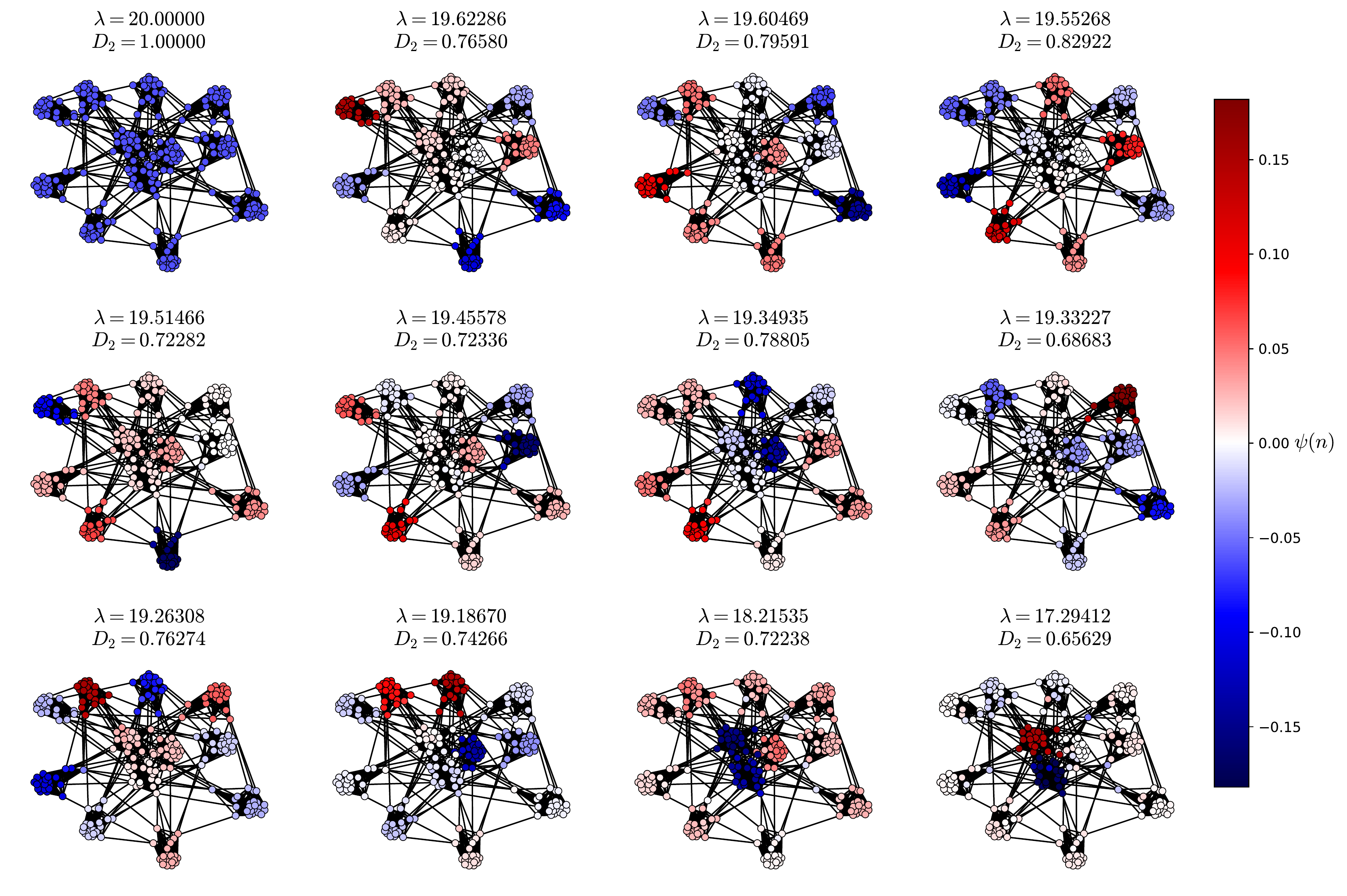}
 \caption{Eigenstate distributions for state in non-perturbative band in the interacting clusterized phase for $k=3$, $d=20$, $N=256$.}
 \label{fig:eigenfunctions}
\end{figure}

Due to the cluster interaction, the eigenfunctions are delocalized between clusters. In order to recognize the delocalization pattern we consider the support of wave functions, corresponding to all states in the non-perturbative band, see Fig.~\ref{fig:eigenfunctions}, for a certain typical realization. Hence in some sense, we could speak about two types
of eigenvalue instantons. Note that the negative eigenvalue instantons
have been recently found in the matrix models~\cite{marino2022new}. 

 For RRG we have the homogeneous state at $\lambda=d$, and all other states in non-perturbative band are orthogonal to it. 
 The rest of the states are different bound states of weakly connected "cluster-particles" and "cluster-holes". There is one state localized at the single cluster and has the structure of the "Particle-hole dipole".




\section{Anderson transition with the diagonal disorder in the $\mu_k$-perturbed RRG 
}\label{Sec:ALT}

In this Section, we consider the simulation of the $\mu_k$-perturbed RRG, supplemented with the i.i.d. random diagonal disorder $\epsilon_i$ of the amplitude $W/2$ of the homogeneous distribution, $|\epsilon_i|<W/2$. 
In the conventional framework, one studies Anderson transition for non-interacting spinless fermions hopping over RRG with connectivity $d=3$ in a diagonal disorder described by Hamiltonian 
\begin{equation}
H=\sum_{i,j}A_{ij}\left(c_i^+c_j+c_ic_j^+\right)+\sum_{i=1}^{ N}\epsilon_i c_i^+c_i,
\label{eq01}
\end{equation}
where the first sum, representing nearest-neighbor hopping between RRG nodes, is written in term of the adjacency matrix, while the second sum, running over all $N$ nodes, represents the potential disorder. 
The pure RRG ensemble undergoes the Anderson localization transition at $W_c=18.16$ for $d=3$~\cite{Luca2014,Kravtsov2018nonergodic,Parisi2019anderson,Tikhonov2021from}. 
For larger $d$ the critical disorder is usually estimated as 
\begin{gather}\label{eq:W_ALT_RRG}
 W_c(d) \simeq d \ln d \ .
\end{gather}

\begin{figure}[h!]
 \centering
 \includegraphics[width=\textwidth]{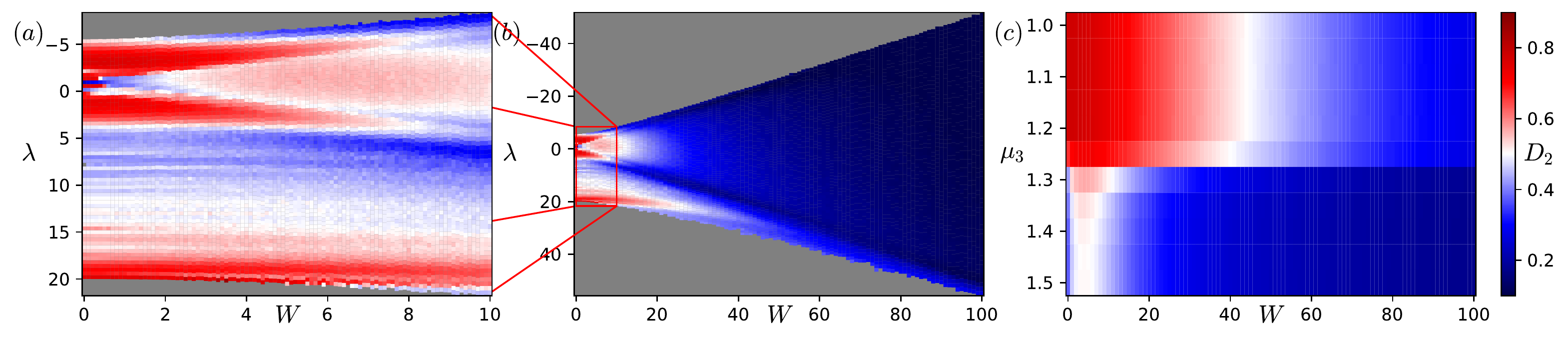}
 \caption{The evolution of energy-resolved fractal dimension for RRG $N=1024$, $d=20$ in the interacting clusterized phase with diagonal disorder $W$, plotted versus (a,~b)~eigenenergies for $\mu_3 = 2$ for different $W$ ranges, as well as 
 (c)~the average fractal dimension versus the chemical potential $\mu_3$. Each point of a color plot is averaged over $20$ (a,~b) or $5$ (c) structural and $5$ (a,~b,~c) disorder realizations.
}
 \label{fig:anderson}
\end{figure}

Now we perturb RRG with a finite chemical potential $\mu_k$ and focus on $k=3$.
The procedure goes as follows: first, we fix the value of $\mu_3$ and the structure of the graph and then investigate the eigenstate localization properties via fractal dimensions over the spectrum for different values of diagonal disorder strength $W$. No full back reaction of the disorder on the graph architecture is considered. 
As soon as chemical potential is strong enough, $\mu_3>\mu_c(3)$, RRG undergoes the clusterization transition. 
We focus on the case of $N=1024$, $d=20$, where the clusterization phase transition occurs approximately at $\mu_3=1.30$.

There are three features found for the case of such combined disorder
\begin{itemize}
 \item The first question concerns the position of the Anderson transition $W_c$
at $(\mu_3,W)$ parameter space. From the 
Fig.~\ref{fig:anderson}(c) one can see that critical value of disorder, where the Anderson localization of all spectrum
at $d=20$ takes place gets significantly reduced in the clustered phase, compared to $W_c\approx 45$, Eq.~\eqref{eq:W_ALT_RRG} in the unclustered phase. 

This reduction of $W_{c}$ is consistent with the structure of the interacting clusters, considered in Sec.~\ref{Sec:semi-Poisson_side_band}. Indeed, in the clustered phase each node in a certain cluster is coupled to $d_0\ll d$ nodes, from other clusters. Thus, effectively the graph degree is reduced, leading, according to the formula~\eqref{eq:W_ALT_RRG}, to the reduction of $W_c$.

\item In addition to the first item, at intermediate disorders, the structure of modes in the perturbative mid-spectrum band is quite peculiar. At small, but finite disorder $W\lesssim 1$ it shows it shows quick localization close to $\lambda=-1$, followed by the maximum in the fractal dimension at intermediate disorder, $W\simeq 10$ and the localization afterwards, $W\gtrsim 15$, see Fig.~\ref{fig:anderson}(a, b).

The origin of this peculiar $W$-dependence is determined by the underlying TENs.
Indeed, at small disorder the mode structure resembles their TEN structure of $W=0$.
Indeed, there are degenerate TENs at $\lambda=-1$, see Sec.~\ref{SubSec:TEN_E=-1}, and the triangle TEN band around it, Sec.~\ref{SubSec:TEN_band}.

Being degenerate at $W=0$, both these types of TENs cannot be distinguished by numerical simulations and, thus, form any superposition within the degenerate manifold.
To separate TEN states, the results for $W=0$ have been calculated with a tiny disorder, of $W=10^{-5}$.
Small disorder, first, lifts these degeneracies and makes wave functions TEN-localized already in the zeroth order in small $W$.
As the number of degenerate TENs is largest at $\lambda=-1$, this part of the spectrum shows the maximal delocalization-to-localization effect at small disorder.

Next, at intermediate disorder values, TEN modes at different energies get hybridized by the diagonal disorder, $W\gtrsim 2$, and then gradually lose their TEN structure, $W\simeq 10$. The competition between TEN- and Anderson localization makes fractal dimension, first, to grow and then decay to zero, when Anderson wins, $W\gtrsim 15$.


\item Finally, the simulation shows, see Fig.~\ref{fig:anderson}, 
that in the clustered phase, the states in the perturbative mid-spectrum band get localized at smaller critical disorder $W_{c,1}$, compared to the ones in the non-perturbative side-band $W_{c,2}>W_{c,1}$. So, there is an intermediate regime $W_{c,1}<W<W_{c,2}$, where the states in the mid-spectrum zone are already localized, while the non-perturbative side-band ones are delocalized, hence there is an effective mobility edge.
This can be also explained both by the initial more localized TEN structure in the mid-spectrum band and by the larger effective degree of the cluster mode graph, $d_{eff} \simeq N_k \mean{d_0}$, see Sec.~\ref{Sec:semi-Poisson_side_band}, with respect to the mid-spectrum.

\end{itemize}


\section{Perturbed RRG from $2$d quantum gravity perspective }\label{Sec:gravity}
In this Section, we focus on the second face of RRG as the tool for the
summation over the 
triangulations of surfaces or therefore over $2$d metrics hence being the discrete model of $2$d quantum gravity.
The summation
over triangulations can be mapped to $d=3$-degree RRG ensemble which finally gets
mimicked by the perturbation theory in Hermitian matrix model with the cubic potential~\cite{Kazakov:1985ea,David:1985et,Kazakov:1985ds}. We shall question if our 
findings concerning the clusterization of perturbed RRG and TENs have the clear-cut 
interpretation in the $2$d quantum gravity.

\subsection{Eigenvalue instantons and anti-instantons at finite N 
and dual gravity configurations}
General $2$d gravity perspective relevant for our study looks as follows.
At large $N$ it is possible to consider the effective $2$d gravity coupled to some matter whose propagator is fixed by the graph Laplacian. For the massless spinless fermions 
at the fluctuating surface the $c=-2$ theory emerges at the criticality~\cite{Kazakov:1985ea,kostov1987random,boulatov1986analytical},
while for the massive spinless fermion the interpolation between $c=0$ and $c=-2$ in the 
large $N$ limit is possible~\cite{gorsky2023flow}. From the $2$d gravity viewpoint
in this study we considered discrete pure gravity interacting with two different matter. There is one type of matter with large mass $m$ which we have integrated out yielding the massive determinant. The $\log \det (A-m^2)= \Tr \log(A-m^2)$ term in the effective action is expanded in inverse powers 
of $m^2$ yielding the chemical potentials for the cycles $\mu_k$. We have truncated
the series and take into account the short cycles only. Since we analyzed the ground state of the partition function with non-vanishing chemical potential we could claim that the back
reaction of the heavy matter on $2$d gravity is taken into account. After deriving the saddle point
configuration we 
consider the massless matter propagating in the background of emerging geometry.

It is worth making the following remark. The matrix
model combinatorics tells that the factor $\frac{1}{N}$ counts the genus of the 
Riemann surface hence at large $N$ we consider the planar RRG ensemble.
Usually the canonical ensemble in $2$d quantum gravity is considered, so that we introduce
the chemical potential for the number of nodes which has the physical interpretation
of $2$d cosmological constant $\beta$. Contrary, in the RRG ensemble the microcanonical
ensemble is considered with the fixed number of nodes. They are related via 
the Laplace transform. 

There is the critical value $\beta_c$ when the partition
function is dominated by large area  (number of nodes) and the continuum limit is available. The
critical exponents can be evaluated from the critical behavior of the partition function
\begin{equation}
 Z(\beta)= \sum_{RRG} \exp(-\beta (\text{Area})) \ .
\end{equation}
It is useful to introduce the variable $g=\exp(-\beta)$, then in the planar limit 
the non-analytic part of the canonical partition function behaves at $g\rightarrow g_c$ as 
\begin{equation}
 Z(g)\propto (g_c-g)^{2-\gamma}
\end{equation}
On the other hand, the leading non-analytic part of microcanonical partition function at fixed number of nodes behaves 
at large area as 
\begin{equation}
 Z_0(A) \propto (\text{Area})^{\gamma-3}
\end{equation}
where the critical exponent $\gamma=\frac{1}{2}$ in the planar limit  can be evaluated via the matrix model or Liouville theory, see \cite{di19952d} for the review.

The focus of our study is the account of the non-perturbative effects which 
in the large $N$ matrix model are
typically suppressed by $\exp(-N)$ factor. Since we consider the numerical
simulations at finite $N$, such non-perturbative instanton which corresponds
to the eigenvalue tunneling in the spectrum can be clearly identified and analyzed.
The most natural interpretation of the eigenvalue instanton in our setting 
is the creation of the baby Universe as initially suggested in~\cite{david1993non}.
Our ideal clusterization corresponds to the creation of the finite number of
non-interacting baby Universes when all degrees of freedom inside them are frozen
and we see just a few peaks in the DOS. If the clusterization is non-ideal
the baby Universes start to interact and the internal modes inside them 
get exited yielding several continuum bands clearly seen in the numerical simulations.

For the odd-$k$ cycles we create the usual clusters while for even cycles they are bipartite.
Hence instead of the eigenvalue instanton we could a bit loosely speak about
the instanton-anti-instanton pair with total vanishing "energy". This situation 
strongly resembles the one considered in~\cite{marino2022new} where the anti-instantons in the matrix models are clearly identified. The bipartite cluster 
can be considered as a "hole" in the graph. It seems that the wormhole gravity
interpretation of the bipartite cluster is also consistent. This can be seen most clearly
if we look at the complimentary graph for the bipartite cluster
when we have the trumpet-like configuration for the bipartite cluster. Remark that the 
clusters corresponding to the negative energy isolated eigenvalues can be induced if we consider negative chemical potential $\mu_3$~\cite{avetisov2018phase} that is different sign of the 
energy in the Hamiltonian of perturbed RRG. In such case we would have anti-instantons
without instantons. However this interpretation certainly requires further analysis.

Note that some care is required in relation between the "RRG matrix model" for 
summation over triangulations with bimodal
matrix elements and the Hermitian matrix ensemble. The latter presumably 
can be interpreted as the holographic dual of $2$d gravity with the 
random Hermitian matrix interpreted as the random Hamiltonian in the 
boundary theory. Such viewpoint has been recently advocated in the 
context of the matrix model 
for the Jackiw-Teitelboim 2d gravity~\cite{saad2019jt}.

In general the matrix model canonical partition function reads as 
\begin{equation}
 Z(t_k) = \int dX \exp\left(\sum_k t_k \Tr X^k \right)
\end{equation}
for potential $V(X)=\sum_k t_k \Tr X^k$ where the integration over the different matrix ensembles can be done. 
Two comments concerning the matrix model approach at large $N$ are in order. First,
the spectral density and all interesting correlators in this limit can be read off from the
spectral curve
\begin{equation}
 y^2(x)= (V'(x))^2 +f(x)
 \label{spectral}
\end{equation}
when the polynomial function $f(x)$ has subleading power compared with the first term. It is 
clear that the genus of the spectral curve depends on the potential and increases with its power.
A slightly different form of the spectral curve reads as
\begin{equation}
 y(x) = V'_{eff}(x)
\end{equation}
where $V_{eff}(x)$ is the effective potential for the eigenvalues which is constant 
at the cuts where the spectral density is supported. For the one-cut solution with
support $[a,b]$ the spectral curve reads as
\begin{equation}
 y(x)=F(x)\sqrt{(x-a)(b-x)}
\end{equation}
with polynomial $F(x)$ called moment function and the condition $F(x^*)=0$
defines the extrema of $V_{eff}(x)$. 

The second point to be mentioned concerns the eigenvalue instantons
that are exact counterparts of the clusters in RRG. The eigenvalue
instanton corresponds to the tunneling of the individual 
eigenvalue from the end of the cut to the nearby extremum $x^*$ of the
$V_{eff}(x)$. The action on the eigenvalue instanton is 
\begin{equation}
 S_{ins}= \int_a^{x*} y(x)dx = V_{eff}(x^*) - V_{eff}(a)
\end{equation}
The instantons are exponentially suppressed at large $N$ as
$\exp(-1/g_s)= \exp(-N)$ where $g_s= \frac{1}{N}$ is the string coupling constant. Multiple eigenvalue instantons can
interact and fill the second non-perturbative band. 

For the pure $3$-degree RRG canonical ensemble the potential in the
Hermitian matrix model is cubic~\cite{Kazakov:1985ds,David:1985et,Kazakov:1985ea}
\begin{equation}
 V_{RRG}(g) = N(-\Tr X^2 + g \Tr X^3)
\end{equation}
where $g= \exp(-\beta)$.
If we add the mass $m$ spinless fermion interacting with random geometry
it induces the $\Tr \log(A-(m^2-3)$ potential in the canonical 
partition function of $3$-degree RRG and the corresponding potential in the
Hermitian matrix model reads as follows~\cite{gorsky2023flow}
\begin{equation}
 V_{RRG+m}(g,m) = N\left(- \Tr X^2 + c\frac{m^2}{g^2} \left(1- \sqrt{1-4g X}\right)^3\right)
\end{equation}
with fixed constant $c$.

To approach our numerical results it would be useful to get carefully 
the similar matrix model
for the exponential random graph for $\mu_k$-perturbed RRG without 
planar approximation. The power of leading term in the potential $\Tr X^k$ in such matrix model certainly 
is determined by the length of the cycles in perturbed RRG but the
careful analysis is required to determine the subleading terms. Our
numerics confirms that the number of bands where the DOS has support
is defined by $k$ is in agreement with expectations.
It would be very interesting to compare the eigenvalue instantons and anti-instantons
in such matrix model with the eigenvalue instantons and anti-instantons in the perturbed
RRG ensemble we have discussed numerically in this study.

\subsection{Gravitational scar states and singular triangulations}
The second natural question concerns the gravity interpretation of the
scar states we have found for perturbed RRG. To start with let us mention
that recently the scar states have been found for probe at the vicinity of the
AdS black hole~\cite{dodelson2022gravitational}. The scars correspond to the classically
stable orbits around the black holes. They are not absolutely stable since 
there is gravitational emission and quantum tunneling which amount to the thermalization at very large times. Hence the authors used the term perturbative scars for these states.
Holographically, these orbits correspond to the states involving double twist operators
in the boundary theory and their weak instability corresponds to the small imaginary parts
in the corresponding Green's functions.

In our framework we have a discrete version of $2$d quantum gravity interacting with matter hence one 
could look for some analogue of the gravitational scars for the probe around the 
particular "local gravitational state". First question 
concerns the "local" analogue of black hole in our case. Fortunately we know what pattern of triangulation 
the TEN-scar states correspond to. The simplest dipole TEN corresponds to the singular
triangulation when $\det A=0$ and therefore we have no inverse matrix. 
The rank of adjacency matrix for this special triangulation decreases
and since the rank of the corresponding matrix is related to the 
genus of the triangulated surface and the number of boundaries (or punctures)
\cite{2006math......8367F} we could claim that TEN corresponds to the 
peculiar singular point on the surface or the hole.
This TEN plays
the role of the local defect of the geometry in the discrete case. 

Instead of the search of the classical stable orbits around black
hole in the continuous case 
we solve the matrix Schr\"odinger
equation for the particle propagating on this graph. The scar state corresponds
to the particle, localized say at dipole TEN pair. The particle, localized at a TEN set, 
can be considered as the discrete analogue of the stable probe orbit around the black hole in~\cite{dodelson2022gravitational}. 
Such gravitational scars can exist in the perturbed RRG in the unclustered 
phase if $\mu_3>\mu_{TEN}(3)$ hence the scars are not related to the clusters
directly. However as we have discussed above, TENs tend to be the precursor and nuclei 
of the emerging ideal cluster which can be seen either at the level of spectrum
or directly in the Monte-Carlo rewiring process. Since such cluster corresponds
to the eigenvalue instanton we could claim that gravitational scars play the role of the
precursor and nuclei for the creation of the baby Universe.

Another attempt to identify holographic scars 
has been performed in~\cite{liska2022holographic}. In that paper, 
it was assumed that the proper representations of the Virasoro algebra
or the Virasoro co-adjoint orbits at the semi-classical level do the 
job. This is parallel to the approach to $2$d gravity via the Liouville
theory and the Teichmuller moduli space. Instead of summation over triangulations
the integration over the moduli space is performed. The relation between
two approaches is well-known in terms of the matrix models since 
there is an equivalence between the large $N$ matrix model with the cubic 
potential, triangulating the surfaces, and the Kontsevich finite $N$ matrix model
in external field, triangulating the moduli space. It would be interesting to recognize TENs
precisely in the language of the moduli space.

To complete this Section let us mention one possible analogy.
The Monte-Carlo process indicates the spin glass-like structure
at the intermediate stages and the very clustered phase has a lot
in common with the one-step replica symmetry breaking. Such underlying
spin-glass pattern implies the presence of multiple metastable states
that provide the playground for the possible false vacuum decay processes.
It is well-known that the false vacuum decay in $1+1$ theory goes
through the formation of the bounce configuration at the Euclidean
space. The radius of the bounce configuration is derived as the
saddle point solution when the pressure term due to 
energy density difference and the surface tension compete.
The boundary of the bounce is the circle trajectory of the 
kink-antikink pair on the Euclidean plane.

As we have mentioned above, TENs are the nuclei of the cluster formation,
hence let us speculate that the TEN dipole serves as analogue of the kink-antikink
pair for the bounce configuration. By definition, the nodes in TEN dipole
interact with the environment identically as expected for kink-antikink pair.
The cluster is built on the top of the TEN dipole and the region inside
the cluster is enriched with the triangles (for $\mu_3\neq 0$) hence 
we have the desired pressure term proportional to $\mu_3$, extending
the analogy. The most subtle point concerns the analogue of the 
tension term which should compete with the pressure to determine
the size of the cluster. Presumably this term is effectively generated 
by the degree conservation rule, since the size of the cluster knows
about the node degree. It would be interesting to check the validity
of this speculation.



\section{Conclusion}\label{Sec:Conclusion}
In this study, we have investigated the phase diagram and localization properties of the $d$-degree RRG ensemble of size $N$, perturbed by the chemical potential $\mu_k$ of the short $k$-cycles, in the $(\mu_k, d, N)$ space. It is found that the DOS exhibits four regimes with different clusterization and localization
patterns. In particular, the phase with ideal clusters gets identified by
highly degenerate spectrum, while at larger degrees there emerges the phase with interacting non-ideal clusters. 
For some values of $(N,k,d)$ the phase of ideal clusters can be absent at all. The clusterization
phase transition at finite $N$ is of the first order and the number of the corresponding cycles serves as the 
order parameter.

We have found the families of the quantum scar states, localized at the fixed values of energy.
The quantum scar states are identified in the non-clustered phase close to the clusterization transition, as well as in the phase of non-ideal interacting clusters. In the non-clustered phase the crossover between the scarless and
scarful phases has been found both numerically and estimated analytically in the $(\mu_3, d)$ 
and $(\mu_4, d)$ parameter planes. 
These scar states are localized at the particular subgraphs of RRG which we identified as 
TEN dipoles and multipoles, however the most general scar states are still to be formulated. 
Presumably, these are related to the particular representations of the group attributed to the graph.
Remarkably, it turns out that the fragmentation and scarring are related phenomena. Namely,
the scars are the precursors and the nuclei for the clusters formation.

Since the RRG ensemble is simultaneously the model for the $2$d quantum gravity, all 
questions discussed in this study can be
rephrased into the gravitational language. It would be very interesting to investigate 
carefully the gravitational meaning of the scar states we have found. At the 
simplest level, we have argued that the scar dipoles correspond to the singular 
triangulations which, on the other hand, correspond to the presence of the 
puncture or the boundary at the surface. Certainly, this issue is more
involved and presumably the scar states can be related to the representation
of the cluster algebras, corresponding to the triangulations of the surfaces.

The adding of the diagonal disorder provides an even more complicated phase structure in the $(\mu_3,W))$ plane. The Anderson localization transition occurs at $W=W_c$ at small $\mu_3$ in the unclustered phase and $W_c$ 
gets reduced by the order of magnitude at $\mu_3>\mu_c(3)$ when the clusterization
phase transition occurs. 
Interestingly, in the later regime in the clusterized phase, there is the range of disorder strength $W_{c,1}<W< W_{c,2}$, where an effective mobility edge gets
emerged: the states in the perturbative mid-spectrum band are already localized, while the ones in the non-perturbative 
side-band still show delocalization. 

There are several questions that deserve further studies. First, it would be interesting to relate our findings to the behaviour of the many-body system in the physical state, modelled by the single-particle dynamics on RRG. The diagonal disorder on RRG is the standard framework for this, while
the chemical potentials for short cycles correspond to account of the higher resonances~\cite{basko2011weak}. 
Combination of the structural and diagonal disorder we have discussed could provide additional viewpoints concerning the tricritical point suggested in~\cite{LN-RP_RRG,LN-RP_WE,LN-RP_dyn}. 

The interplay of our findings with matrix models is two-fold. First, we have a perfect 
playground for the numerical investigation of the non-perturbative phenomena at finite $N$ in RRG,
considered as the version of the matrix model. The interaction of the eigenvalue instantons in the matrix models is precisely mapped into interaction of clusters in clusterized phase of RRG and we hope to discuss these issues in more details in a separate study. In particular, it would be interesting to investigate numerically the properties of the "holes" in the graph, that correspond to the anti-instantons in the matrix models. We plan to investigate 
the case of negative chemical potentials that induce the formation of eigenvalue anti-instantons-holes 
elsewhere.

Another issue which certainly is worth to investigate concerns the clusterization transition in the $2$d quantum gravity interacting with massive matter as the function of mass. 
To fully clarify this issue, it is desirable 
to perform the numerical simulations for the massive particle on RRG, that is
RRG with the measure $\Tr\log (A(G) + m^2)$ for graph $G$ without the expansion of the 
determinant at large $m^2$ which produces the chemical potentials for the short cycles. 
Instead of solution of the matrix model corresponding to this setup found in~\cite{gorsky2023flow} 
beyond the planar limit we could perform numerical simulations with this measure for the finite $N$
RRG ensemble at arbitrary $m$.

We could also try to link two faces of perturbed RRG together. That is 
$2$d quantum gravity supplemented with the particular perturbation 
and the model of Hilbert space of the 
interacting many-body system, supplemented by the chemical potentials
for higher resonances. In this language, the critical phenomena 
in the Hilbert space acquire the gravitational flavor. 
The simplest question concerns the real space interpretation
of the critical exponent of the RRG partition function in 
the thermodynamic limit.
Analysis 
of the fractal dimensions of eigenfunctions 
of the probe at RRG is the standard observable which gets translated
into the behavior in the physical space. However, more complicated 
correlators of vertex operators can be discussed in the $2$d quantum gravity
and their properties could provide the additional insights concerning
the properties of the system in the physical space.
Such perspective 
has been discussed for the Dirac operator in the gauge field 
with the diagonal disorder in~\cite{kogan1996liouville},
where the fractal dimension of the zero modes was linked with the conformal dimensions 
of particular operators in the Liouville theory. It would be also 
interesting to interpret the genus, counting in RRG in the physical
many-body system.

We have observed that for $\mu_k$-perturbed RRG for even-$k$ cycles the bipartiteness has been restored in the clustered phase and the spectrum of adjacency matrix 
is symmetric with respect to $\lambda=0$. This suggests the interesting analogy with
the chirality in the fermionic systems and the corresponding symmetry of the Dirac 
operator. Having in mind some well established properties of the Dirac operator spectrum
in QCD, we could look for the similar results, like Casher-Banks relation~\cite{banks1980chiral}, in the perturbed RRG context. In the instanton liquid model~\cite{schafer1998instantons} the restoration of the chiral symmetry at the deconfinement
phase transition presumably is accomplished with the formation of the instanton 
clusters~\cite{schafer1995chiral}. This resembles our restoration of the 
bipartiteness via clusterization. We plan to discuss this analogy elsewhere.

It would be also interesting to investigate the effects of the clusterization in the partially disordered RRG ensemble~\cite{valba2022mobility}. In that case, there is the critical ratio of the number of dirty and clean nodes,
when the delocalized part of the spectrum exists even at arbitrarily
large value of the partial disorder amplitude $W$. The clusterization and TEN modes are expected to influence this behavior.


\subsection*{Acknowledgements}
I. M. K. acknowledges the support by the European Research Council under the European Union’s Seventh Framework Program Synergy No. ERC-2018-SyG HERO-810451. The work of A.G. was supported by grant of Basis Foundation 20-1-1-23-1.
A.G. thanks Nordita for the hospitality and support.

\clearpage
\begin{appendix}
\section{DOS for the different cycles}\label{App:4-6_cycles}
In this Appendix we collect the plots of the DOS for the $\mu_k$-deformed RRG
which were used in the derivation of the numerical critical curve $\mu_c(k,d,N)$.


\subsection{4-cycles}
\begin{figure}[h]
\centering
\includegraphics[width=0.95\textwidth]{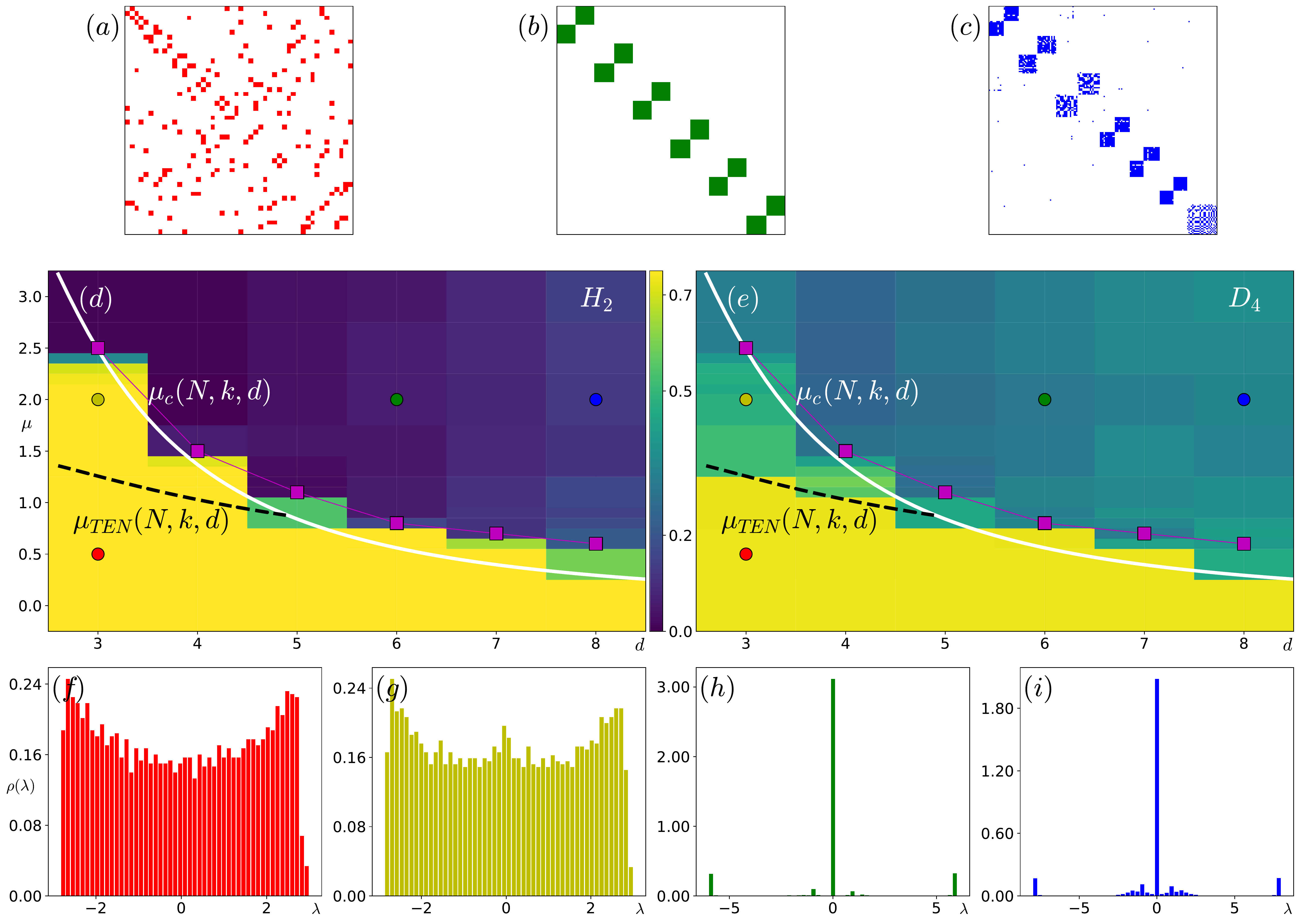}
\caption{Phase diagram in the plane ``chemical potential~--~vertex degree'' $(\mu_k,d)$ for finite-size RRG graphs for $N=256$ and $k=4$-cycles, similar to Fig.~\ref{fig:fig1_phase_diag}. (a-c)~The cluster structure of RRG in (a)~unclustered; (b)~ideally clustered, and (c)~interacting clustered phases. In the case of even $k$ the graph is bipartite in the clustered phase. (d-e)~Phase diagram, with drastic changes (d)~in the density of states (DOS) via the Hellinger distance with respect to the ideal cluster, showing the clustering transition (purple squares), and (e)~in the higher-order fractal dimension $D_4$, sensitive to the scar states, given by the topologically equivalent nodes (TEN).
Panels~(f-i) show the averaged DOS in each of the $4$ phases:
(f)~unclustered, (g)~TEN-scarred unclustered, (h)~ideally clustered, and (i)~interacting clustered phases. 
The colors of the solid circles in the panels (d, e), marking each of $4$ phases, correspond to the colors of the blocks in (a-c) and the DOS in (f-i). Solid white $\mu_c(N,k,d)$, Eq.~\eqref{eq:mu_c(N,k,d)}, and dashed black $\mu_{TEN}(N,k,d)$, Eq.~\eqref{eq:mu_TEN}, lines show analytical estimates for the transition lines between the above phases.}
\label{fig:pd_4cyc} 
\end{figure}


\clearpage
\subsection{5-cycles}

\begin{figure}[h]
\centering
\includegraphics[width=0.95\textwidth]{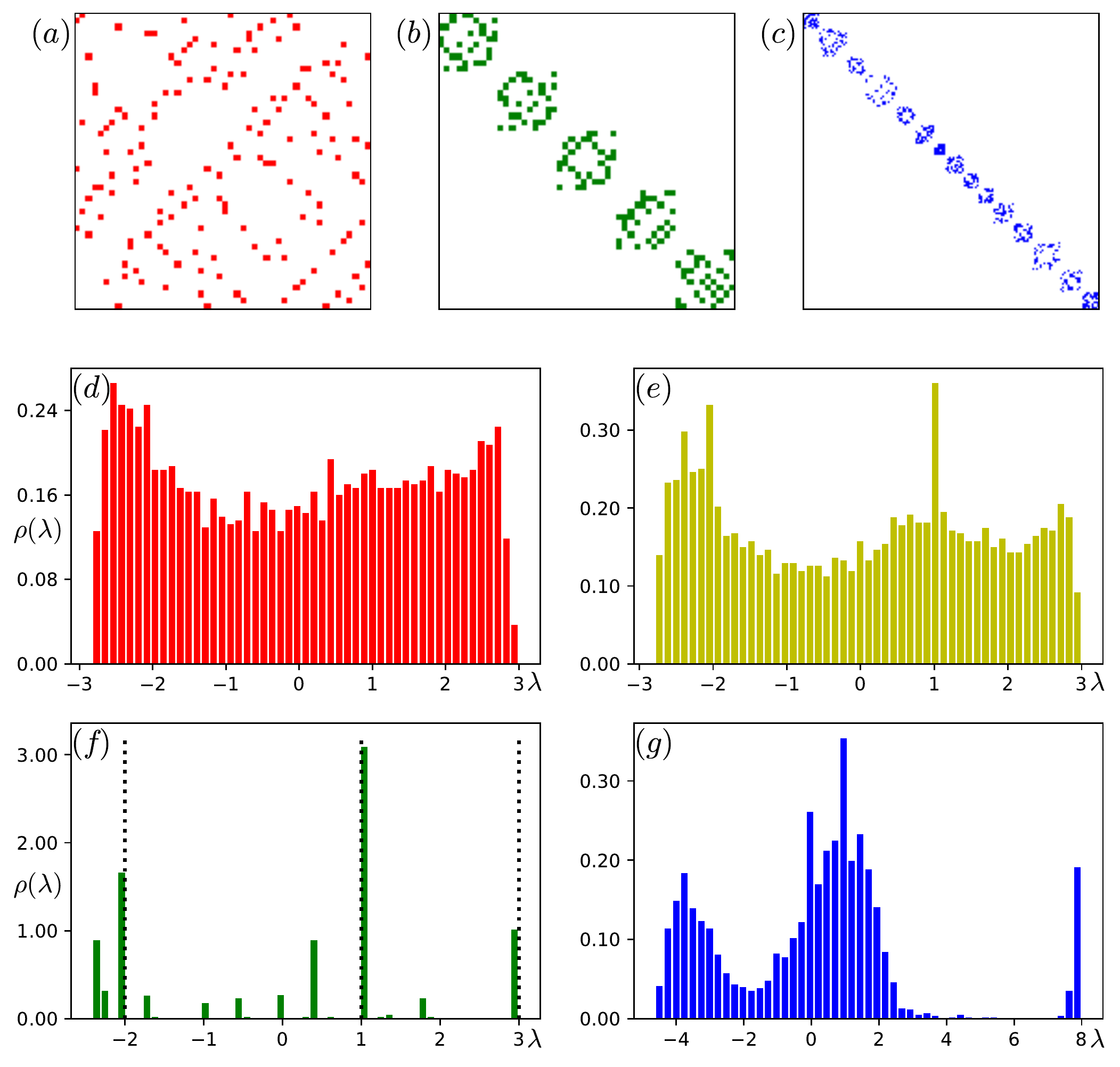}
\caption{The cluster structure and the averaged DOS for finite-size RRG graphs for $N=256$ and $k=5$-cycles, similar to Fig.~\ref{fig:fig1_phase_diag}. (a-c)~The cluster structure of RRG in (a)~unclustered; (b)~ideally clustered, and (c)~interacting clustered phases. 
Panels~(d-g) show the averaged DOS in each of the $4$ phases:
(d)~unclustered, (e)~TEN-scarred unclustered, (f)~ideally clustered, and (g)~interacting clustered phases.}
\label{fig:pd_5cyc} 
\end{figure}

\clearpage
\subsection{6-cycles}

\begin{figure}[h]
\centering
\includegraphics[width=0.95\textwidth]{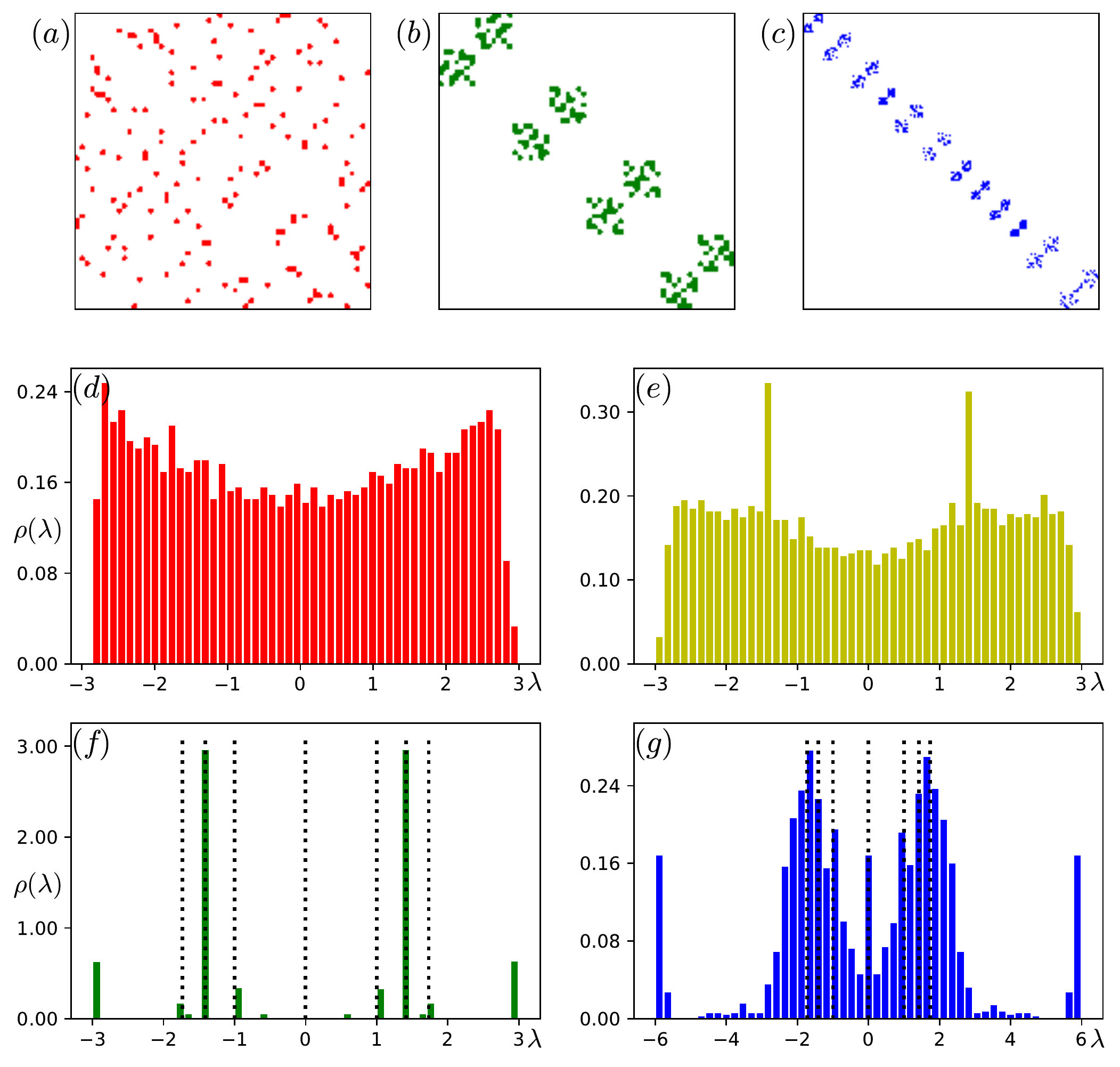}
\caption{The cluster structure and the averaged DOS for finite-size RRG graphs for $N=256$ and $k=6$-cycles, similar to Fig.~\ref{fig:fig1_phase_diag}. (a-c)~The cluster structure of RRG in (a)~unclustered; (b)~ideally clustered, and (c)~interacting clustered phases. In the case of even $k$ the graph is bipartite. 
Panels~(d-g) show the averaged DOS in each of the $4$ phases:
(d)~unclustered, (e)~TEN-scarred unclustered, (f)~ideally clustered, and (g)~interacting clustered phases. The vertical dashed lines correspond to (f)~the non-trivial eigenstates of the ideal cluster, (g)~to $\pm\sqrt{n}$ TEN-scar states, discussed in Sec.~\ref{SubSec:TEN_even_sqrt_n}. }
\label{fig:pd_6cyc} 
\end{figure}

\clearpage


\clearpage


\section{Example of more generic scar complex}\label{App:TEN_complexes}
\begin{figure}[h]
\centering
\begin{tikzpicture}[roundnode/.style={circle,minimum size=0.25cm},squarednode/.style={rectangle,minimum size=0.75cm}]

\node[squarednode,fill=blue] at (2,0) (b1) {$b_1$}; 
\node[squarednode,fill=blue] at (2,2) (b2) {$b_2$};
\node[squarednode,fill=blue] at (4,0) (b3) {$b_3$};
\node[squarednode,fill=blue] at (4,2) (b4) {$b_4$}; 
\node[squarednode,fill=blue] at (3,3.5) (b5) {$b_5$};

\draw[-] (b1) -- (b2);
\draw[-] (b1) -- (b3);
\draw[-] (b1) -- (b4);
\draw[-] (b1) -- (b5);
\draw[-] (b2) -- (b3);
\draw[-] (b2) -- (b4);
\draw[-] (b2) -- (b5);
\draw[-] (b3) -- (b4);
\draw[-] (b3) -- (b5);
\draw[-] (b4) -- (b5);

\node[roundnode,fill=green] at (0,1) (g1) {$g_1$};

\draw[-] (b1) -- (g1);
\draw[-] (b2) -- (g1);
\draw[-] (b3) -- (g1);
\draw[-] (b4) -- (g1);
\draw[-] (b5) -- (g1);

\node[roundnode,fill=yellow] at (6,1) (y1) {$y_1$};
\node[roundnode,fill=yellow] at (6,3) (y2) {$y_2$};

\draw[-] (b1) -- (y1);
\draw[-] (b2) -- (y1);
\draw[-] (b3) -- (y1);
\draw[-] (b4) -- (y1);
\draw[-] (b5) -- (y2);

\node[squarednode,fill=magenta] at (8,1) (m1) {$m_1$};
\node[squarednode,fill=magenta] at (8,3) (m2) {$m_2$};

\draw[-] (m1) -- (m2);
\draw[-] (m1) -- (y1);
\draw[-] (m2) -- (y2);

\node[roundnode,fill=red] at (10,2) (r1) {$r_1$};

\draw[-] (m1) -- (r1);
\draw[-] (m2) -- (r1);

\end{tikzpicture} 
\caption{Example. Eigenvector is nonzero on squared nodes. Green and red nodes are denoted all common neighbors for all connected squared nodes. Eigenvector on yellow nodes equals zero, but these nodes are connected to a nonzero subsets, and with their support, all squared nodes form TEN.}
\label{fig:example} 
\end{figure}
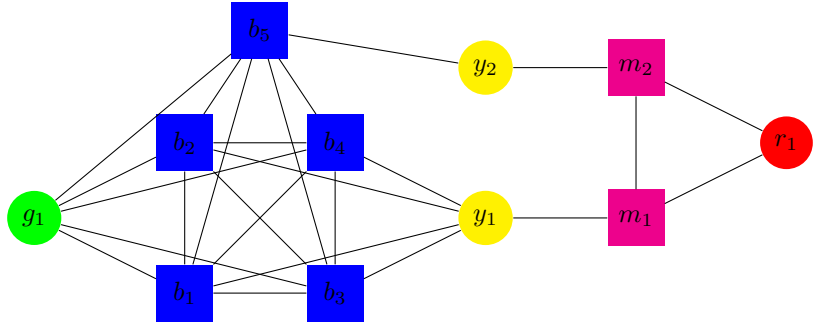

In this Appendix we comment on more general TEN
complex - subgraph which supports non-thermalizing state.
In fact, we have no desired generic definition yet and restrict ourselves by an example.
Let's analyze the example which shows the possible non-trivial structure of TEN (see Fig.~\ref{fig:example}). The equation for nodes, where the eigenvector is localized, reads as
\begin{equation}
	\begin{cases}
	g_1 + b_2 + b_3 + b_4 + b_5 + y_1 = \lambda b_1\\
	g_1 + b_1 + b_3 + b_4 + b_5 + y_1 = \lambda b_2\\
	g_1 + b_1 + b_2 + b_4 + b_5 + y_1 = \lambda b_3\\
	g_1 + b_1 + b_2 + b_3 + b_5 + y_1 = \lambda b_4\\
	g_1 + b_1 + b_2 + b_3 + b_4 + y_2 = \lambda b_5\\
	r_1 + m_1 + m_2 + y_1 = \lambda m_1\\
	r_1 + m_1 + m_2 + y_2 = \lambda m_2\\
	\end{cases}
 \label{eq:example}
\end{equation}
Firstly, nodes $b_1, b_2, b_3, b_4$ form the topological equivalent quadruple. The first four equations in (\ref{eq:example}) give three degenerate states with $\lambda=-1$, with the only condition $\sum_{l=1}^4 b_l = 0$. Secondly, nodes $b_1, b_2, b_3, b_4, b_5, m_1, m_2$ form a more complex system that gives one more degenerate state with $\lambda=-1$. The example shows that nodes do not have to be connected to each other to be part of TEN. 

We present conditions, that might be part of generic definition of complex TENs.
\begin{itemize}
\item TEN cluster shouldn't have external nodes, coupled with the only internal cluster node. For vertex set $I=\{i_1,. . . i_k\}$ general TEN complex must obey
\begin{equation}\label{eq:A_Im}
\sum_{m\in I} A_{Im} + 
\sum_{\substack{m \notin I \\ A_{Im}>1}} A_{Im} = \sum_{i \in I} d_i 
\ . 
\end{equation}
\item Internal structure of TEN subgraph must lead to same eigenvalue for different connected components.
\item Topological equivalence to external nodes. The rank of sub-matrix of the adjacency matrix with elements corresponding to edges from TEN nodes to external nodes must be less than number of nodes in TEN cluster, $\mathrm{rank}\left(A_{Im}\right)<k$, $m \notin I$.
\end{itemize}

\clearpage


\end{appendix}
\bibliographystyle{unsrt}
\bibliography{triangle}

\end{document}